\documentclass[twocolumn]{aastex701}

\shorttitle{Data-driven RMHD Models of Emerging Active Region}
\shortauthors{F. Chen}
\usepackage{amsmath}
\usepackage{color}
\newcommand{\sectref}[1]{Section\,\ref{#1}}

\newcommand{\figref}[1]{Figure\,\ref{#1}}
\newcommand{\figsref}[1]{Figures\,\ref{#1}}
\newcommand{\tabref}[1]{Table\,\ref{#1}}
\newcommand{\equref}[1]{Equation\,(\ref{#1})}

\begin{document}

\title{Data-driven Radiative Magnetohydrodynamics Simulations with the MURaM Code: \\
Coronal Heating and Dynamics in an Emerging Active Region}

\correspondingauthor{Feng Chen}
\email{chenfeng@nju.edu.cn}

\author[0000-0002-1963-5319]{Feng Chen}
\email{chenfeng@nju.edu.cn}
\affiliation{School of Astronomy and Space Science, Nanjing University, Nanjing 210023, China}
\affiliation{Key Laboratory of Modern Astronomy and Astrophysics (Nanjing University), Ministry of Education, Nanjing 210023, China}

\begin{abstract}
We present the application of the data-driven branch of the MURaM code, which follows the evolution of the active region 11640 over 4 days starting from 2012 December 30 at 12:00 UT and reproduces many key coronal extreme-ultraviolet (EUV) emission features seen in remote sensing observations. Radiative magnetohydrodynamic (MHD) simulations that account for sophisticated energy transport processes, such as those in the real corona, have been extended with the ability to use observations as time-dependent boundaries such that the models follow the evolution of actual active regions. This opens the possibility of a one-to-one model of a target region over an extensive time period. We use a hybrid strategy that combines fast-evolving idealized zero-$\beta$ models that capture the evolution of the large-scale active region magnetic field over a long time period and sophisticated radiative MHD models for a shorter time period of interest. The synthesized EUV images illustrate the formation of coronal loops that connect the two sunspots or fan out to the domain boundary. The model reveals in three-dimensional space fine structure in the coronal heating and plasma properties, which are usually concealed behind the EUV observables. The volumetric heating rate in bright coronal loops is proportional to $\mathbf{B}^{2}$. The emission-measure-weighted line-of-sight velocity, which represents the Doppler shift of a spectral line forming in a certain temperature range, reveals vigorous dynamics in plasma at different temperatures and ubiquitous MHD waves, as expected in the real solar corona.
\end{abstract}

\keywords{Radiative magnetohydrodynamics (2009), Magnetohydrodynamical simulations(1966), Solar magnetic flux emergence (2000), Solar corona (1483), Solar active regions(1974), Solar extreme ultraviolet emission (1493)}

\section{Introduction} \label{sec:intro}
The Sun has been observed in great detail. However, the magnetic structures and dynamics of the three-dimensional (3D) volume of the solar atmosphere, which are believed to be key to understanding almost all solar activity, remain untouchable through direct measurements. Therefore, a comparative investigation that combines observations and sophisticated models, especially those that are able to reproduce the conditions of the real Sun and are consistent with the strict observations, has become an edge-cutting approach to address the remaining problems, such as the formation and eruption of solar magnetic field structures and the heating of the outer atmosphere.

Constructing model based on the observed magnetic field is an approach that has been extensively used, particularly in the studies of solar eruption. This approach is often extended to evolving models, which are commonly known as data-driven models that are driven by a time-dependent boundary that follows the observed magnetic field in the photosphere. Most of the models focus strongly on the magnetic field, with a tradeoff on a more accurate description of plasma thermodynamics. For example, the magneto-frictional method or zero-$\beta$ assumption, which largely omits the role of plasma, has been used in modeling the formation and eruption of magnetic structures by \citet{Pomoell+al:2019,GuoYang+al:2019,ZhongZe+al:2021,Lumme+al:2022}, to name a few. Models that solve the full magnetohydrodynamic (MHD) equations \citep[see e.g., ][]{Inoue+al:2018,Hayashi+al:2018,Jin+al:2018,Kaneko+al:2021,Inoue+al:2022} consider the interaction of the plasma and magnetic field. Research on solar eruptions is not the topic of this paper; a much more comprehensive discussion on the application of data-driven MHD simulations can be found in recent reviews in that field \citet{JiangChaowei+al:2022,JiangChaowei:2024,Schmieder+al:2024}. The community has been aware of the necessity of improving the energy equation that governs the plasma thermodynamics, and very recently, such models were presented by \citet{Afanasyev+al:2023,GuoJinhan+al:2023,GuoJinhan+al:2024}, and \citet{Fan+al:2024}.

It is equally challenging to model a noneruptive active region and reproduce fundamental loop structures that are brilliantly observed in the active region corona. In such models, the delicate balance between coronal heating and the cooling through optically thin radiation and thermal conduction is crucial, as demonstrated in classical works \citep{RTV}. Modern 3D realistic coronal models that have emerged in the last 20 years have demonstrated that magnetic field braiding by photospheric granulation self-consistently provides a sufficient upward energy flux than can heat the corona to more than one million K and give rise to coronal loops  \citep{Gudiksen+Nordlund:2005a,Gudiksen+Nordlund:2005b,Bingert+Peter:2011,Hansteen+al:2015,Rempel:2017}.

Although never emphasized, some of these coronal models have employed setups on the basis of observations. For example, \citet{Gudiksen+Nordlund:2005a} used an observed magnetogram to construct the initial condition of the magnetic field, which was braided by a velocity field that mimicked solar granulation. \citet{Bingert+Peter:2011} utilized a similar method and imposed an enhanced network magnetic field in the quiet Sun area. These models are compared with observations in terms of general coronal plasma properties, instead of those in the particular active region that is used to construct the simulation. A more detailed comparative investigation was performed by \citet{Bourdin+al:2013}. In this model, not only is the observed magnetogram adapted to set the initial condition, but a time series of the photospheric velocity field is also used to drive (mostly to braid) the magnetic field. The model reproduces a few loops, albeit not all, that demonstrate Doppler velocities consistent with those obtained from a spectroscopic observation of this particular region. They also reported that the height of the loops in the model is consistent with the stereoscopic observation of the real loops in the target active region. More recently, a data-driven coronal model that was designed to fully cover an extensive active region was reported by \citet{Warnecke+Peter:2019}. The model was driven by a time series of observed magentograms over approximately 4 hours and an artificial velocity field, as implemented by \citet{Gudiksen+Nordlund:2005a}. The synthesized EUV images illustrate loop structures of various lengths and shapes as that particular active region. Global-scale models also employ observations to constrain the model and reproduce coronal structures that can be observed during total solar eclipses \citep{Mikic+al:2018,Downs+al:2025} or focus on a localized active region \citep{ShiTong+al:2022,ShiTong+al:2024}. A major difference is that these global models usually introduce an energy flux based on, for example, alfv\'en wave fluxes to heat the corona instead of self-consistent generation through the interaction of photospheric flows and the magnetic field.

In this paper, we present a method to conduct data-driven radiative MHD simulations for substantially evolving active regions that extends the scope of previous works that have focused mostly on a stable stage. The MURaM code with the corona extension \citep{Rempel:2017} can cope with modeling the general properties of the solar corona across a wide temperature range \citep{Malanushenko+al:2022,Chen+al:2022,Lu+al:2024,Chen+al:2025}, as well as solar eruptions \citep{Cheung+al:2019,Chen+al:2023L,Rempel+al:2023}. These models were designed to match a particular active region or eruption event. Such a task requires the ability to drive the simulation with a time-dependent boundary. As the first paper in this series, \citet{Chen+al:2023} described the basic method, including the governing equations and the implementation of the boundary, and demonstrated that this method was validated by reproducing a ground truth simulation \citep{Cheung+al:2019}.

As a second paper in this series, the purpose is to describe the method of applying the data-driven MURaM code to an actual active region, present the basic properties of the model active region, compare those with observations of this particular region, and illustrate how the model may change with different numerical setups. The target for this study is a calm (meaning not flare-productive) active region (AR) 11640, whose emergence is captured across the solar disk. The results may have interesting applications in several aspects, such as
\begin{itemize}
\item To reveal the evolution of the coronal heating in the 3D space over the course of active region emergence and how it gives rise to the EUV features as observed,
\item To serve as a benchmark of noneruptive emerging active regions, which is a key task of an international collaboration on 3D data-driven models of active region coronae\footnote{Data-driven 3D Modeling of Evolving and Eruptive Solar Active Region Coronae, led by G. Chintzoglou \& M. Wheatland \\https://teams.issibern.ch/solaractiveregioncoronae},
\item To provide a sample of a realistic active region magnetic field and plasma, where new methods for coronal magnetic field measurement \citep{YangZihao+al:2020,YangZihao+al:2024,ChenYajie+al:2021} may be tested.
\end{itemize}
The rest of the paper is organized as follows. We describe in \sectref{sec:method} the method of computing the electric field from the observed photospheric magnetic field and the hybrid modeling strategy. The results are presented in \sectref{sec:result}, including a comparison between model-synthesized and actual observations of the target region, the relationship between the underlying heating rate and the apparent observables, and the analysis of the control experiments. We discuss the aspects in which the current model can be further improved and conclude in \sectref{sec:conclusion}.

\section{Methods}\label{sec:method}

\begin{deluxetable*}{lrcccc}
\digitalasset
\tablewidth{0pt}
\tablecaption{Summary of simulation cases\label{tab:list}}
\tablehead{
\colhead{Case Name} & \colhead{Grid} & \colhead{Spacing} & \colhead{$\Omega$ in $\nabla_{h}\cdot E_{h}$} & \colhead{Time started} & \colhead{Time evolved}  \\
\colhead{} & \colhead{$N_{x}{\times}Nz$} & \colhead{$\Delta x$[km]$\times\Delta z$[km]} & \colhead{[s$^{-1}$]} &  &
}
\startdata
Bevo\_$\Omega$0 & $256\times1152$ & $576\times64$  & 0                          &Day 0.5 & $>3.5$ days \\
Bevo\_$\Omega$3 & $256\times1152$ & $576\times64$  & $-3\times10^{-6}$ &Day 0.5 & $>3.5$ days \\
Bevo\_$\Omega$5 & $256\times1152$ & $576\times64$  & $-5\times10^{-6}$ &Day 0.5 & $>3.5$ days \\
\hline
D1\_$\Omega$0    & $512\times1152$ & $288\times64$  & 0                           &Day 1 &2.9 hr \\
D2\_$\Omega$0    & $512\times1152$ & $288\times64$  & 0                           &Day 2 &2.9 hr \\
D3\_$\Omega$0    & $512\times1152$ & $288\times64$  & 0                           &Day 3 &2.9 hr \\
D4\_$\Omega$0    & $512\times1152$ & $288\times64$  & 0                           &Day 4 &2.9 hr \\
\hline
D1\_$\Omega$3    & $512\times1152$ & $288\times64$  & $-3\times10^{-6}$  &Day 1 &2.9 hr \\
D2\_$\Omega$3    & $512\times1152$ & $288\times64$  & $-3\times10^{-6}$  &Day 2 &2.9 hr \\
D1\_$\Omega$5    & $512\times1152$ & $288\times64$  & $-5\times10^{-6}$  &Day 1 &2.9 hr \\
D2\_$\Omega$5    & $512\times1152$ & $288\times64$  & $-5\times10^{-6}$  &Day 2 &2.9 hr \\
\hline
D2\_$\Omega$0\_High & $1024\times1152$ & $144\times64$  & 0                           &Day 2+1 hr &2 hr \\
D2\_$\Omega$3\_High & $1024\times1152$ & $144\times64$  & $-3\times10^{-6}$  &Day 2+1 hr &2 hr \\
D2\_$\Omega$3\_Uniform & $512\times1152$ & $288\times64$  & $-3\times10^{-6}$  &Day 2 &2.9 hr \\
D2\_$\Omega$3\_Upflow & $512\times1152$ & $288\times64$  & $-3\times10^{-6}$  &Day 2 &2.9 hr \\
\enddata
\tablecomments{
All the models use $N_{x}=N_{y}$ and $\Delta x=\Delta y$.}
\end{deluxetable*}

\begin{figure*}[ht!]
\center
\includegraphics[width=18cm]{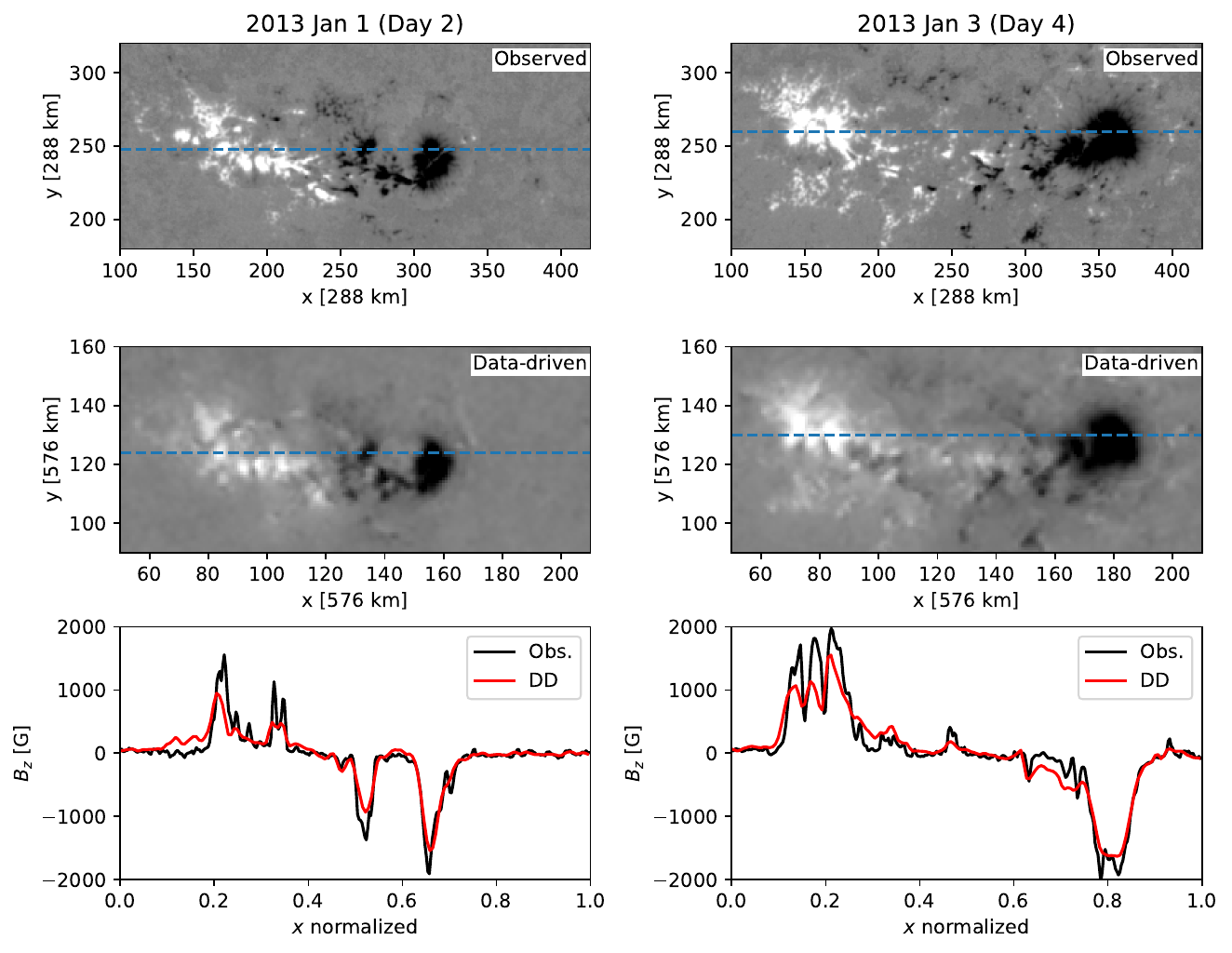}
\caption{Comparison between the observed vertical magnetic field and the vertical magnetic field ($B_{z}$) in the zero-$\beta$ model. The field of view covers the main sunspots in the active region. The top row presents the observed magnetic field with grayscale saturation at $\pm 1000$\,G. The middle row shows $B_{z}$ in the first layer in the computational domain. The bottom row compares the observed $B_{z}$ and that in the data-driven model along the blue dashed lines, as marked in the upper two rows.
\label{fig:boundary}}
\end{figure*}

\subsection{Data-driven Boundary Condition}
The implementation of a time-dependent data-driven boundary for the MURaM code has been described in detail in our previous paper \citep{Chen+al:2023}. The horizontal electric field $\mathbf{E}_{h}=(E_{x}, E_{y})$ at the bottom boundary evolves the magnetic field in the first layer of the domain. When we apply this method to an observed emerging active region, we follow the strategy described by \citet{Cheung+DeRosa:2012}. We use the HMI \citep{HMI,Hoeksema+al:2014} vector magnetic field with a cadence of 720\,s. This cadence works well to capture long-time evolution for both noneruptive active region, as shown in this study, and the flare-productive active region, as shown in \citet{Chen:2026}. Higher-cadence HMI data may provide a better record of the dynamic evolution of the photospheric magnetic field; however, there is a potential risk of spurious energy injections due to the increased noise.

In our study, we use only the radial component and refer to this field as the vertical magnetic field $B_z$, as we use a Cartesian domain. The curl of the horizontal field is constrained by 
\begin{equation}\label{equ:curle}
\nabla_{h}\times \mathbf{E}_{h} = -\frac{\Delta B_{z}}{\Delta t},
\end{equation}
where $\Delta B_{z}$ is the change in $B_{z}$ between two consecutive snapshots with a time difference of $\Delta t$. As demonstrated by \citet{Cheung+DeRosa:2012}, this constraint alone is not sufficient to calculate the two components of the horizontal electric field. An additional constraint can be given by defining the divergence of the horizontal electric field. $\nabla_{h}\cdot \mathbf{E}_{h}$ can be reformulated as
\begin{align}
\nonumber
\nabla_{h}\cdot \mathbf{E}_{h} &=  \frac{\partial}{\partial x}\left(v_{z}B_{y}-v_{y}B_{z}\right)+\frac{\partial}{\partial y}\left(v_{x}B_{z} - v_{z}B_{x}\right) \\
\nonumber
&= \left(\nabla_{h}\times\mathbf{B}_{h}\right)v_{z} - \left(\nabla_{h}\times\mathbf{v}_{h}\right)B_{z} \\ 
&- \left(\mathbf{B}_{h}\times\nabla_{h}\right)v_{z} + \left(\mathbf{v}_{h}\times\nabla_{h}\right)B_{z},
\end{align}
which reveals that the right-hand side contains terms related to the vertical electric current and vertical vorticity. This motivates assigning values of $\nabla_{h}\cdot \mathbf{E}_{h}$ by $B_{z}$ multiplied by a free parameter that represents the rotation of the horizontal velocity field, as suggested by \citet{Cheung+DeRosa:2012}. An alternative option is used by \citet{Cheung+al:2015} in a model of a small-scale rotation jet, as well as by \citet{Fan+al:2024} in an active-region scale model of a large solar eruption. The value of $\nabla_{h}\cdot \mathbf{E}_{h}$ is determined by  $\nabla_{h}\times\mathbf{B_{h}}$ multiplied by a free parameter that represents an emerging motion. With the information provided by the observed magnetic field, it is still impractical to precisely determine the values of $\nabla_{h}\cdot \mathbf{E}_{h}$, and the options above represent a parameterization of the twist added to the magnetic field. 

In this study, we employ the same approach as \citet{Cheung+DeRosa:2012} and $E_{h}$ follows
\begin{equation}\label{equ:dive}
\nabla_{h}\cdot \mathbf{E}_{h} = -\Omega B_{z}.
\end{equation}
The case with $\Omega=0$ leads to a vanishing right-hand side and represents a state with a minimum energy. We note that $\Omega=0$ does not indicate a potential magnetic field. The active region develops free magnetic energy as the sunspots emerge and move in the photosphere. We also use $\Omega=-3\times10^{-6}$ and $-5\times10^{-6}$\,s$^{-1}$. Larger absolute values $\Omega$ represent a stronger rotation of the horizontal velocity field and eventually give rise to a stronger twisted horizontal magnetic field, whereas the vertical magnetic field still follows the observation constraint. 

The sign of $\Omega$ is chosen according to AIA observations of the coronal loops above this active region. Active region 11640 is a very calm active region, in the sense that no major flares occur during more than 4 days of evolution. This means that the amount of free magnetic energy in this region is not abundant. As shown later in the paper, the run cases with $\Omega=0$ can reproduce many fundamental properties of the observed active region. In this study, we use a relatively small $\Omega$ to match the condition of this active region. A nonvanishing $\Omega$ causes an additional shift in the position of the coronal loops, as well as some distortion of their shape, because an extra nonpotential field is injected into the corona. The same trend can be seen in the large loop arches from the AIA images of the active region, which allows us to determine the sign of $\Omega$.

In this study, we conduct simulations with different resolutions. The $700\times290$ array of $B_{z}$ is padded with zero values to a $768\times768$ array. The padded magnetogram is then interpolated linearly into the mesh of corresponding resolution. The electric field is computed by solving \equref{equ:curle} and \equref{equ:dive} via the fast Fourier transform method.

\subsection{Model Strategy}
We employ a two-stage hybrid model that comprises a zero-$\beta$ MHD model, which provides an evolution of the coronal magnetic field in the target active region over the long course of flux emergence, and several radiative MHD models, which provide a more realistic calculation of the plasma properties at short time periods of interest. In a general sense, the basic idea of hybrid models, which usually combines a more simplified but less computationally demanding model and a more self-consistent and sophisticated MHD model, has been employed in many previous studies when covering long-term evolution and resolving detailed dynamics in a short period are both important. For example, \citet{Amari+al:2003I,Amari+al:2003II} combined time series of static magnetic field extrapolations with evolving MHD models. \citet{,Amari+al:2014,Afanasyev+al:2023,GuoJinhan+al:2024} used snapshots of the magnetic field from, for example, an extrapolation or magneto-frictional model as the initial condition for full MHD models. Our implementation of the hybrid model strategy in the data-driven MURaM simulations is as follows.

\subsection{Zero-$\beta$ MHD Model}\label{sec:zb}
The zero-$\beta$ assumption omits all plasma forces and gravity. We solve the momentum and induction equation as described in \citet{Chen+al:2023}, and the right-hand side of the momentum equation is governed by the Lorentz force. A static and height-dependent background density profile is set in this zero-$\beta$ model. A modification of the uniform background density in the test case of \citet{Chen+al:2023}, which is valid for a thin slab near the photosphere, is that we adapt the density stratification from the photosphere to the corona from an already-done quiet Sun MURaM simulation. The quiet Sun simulation extends from the uppermost convection zone to approximately 100\,Mm in the corona, which is similar to that shown in \citet{Chen+al:2022} but with a small horizontal domain extension. The small-scale dynamo in the convection zone generates a mix-polarity small-scale magnetic field that permeates through the whole atmosphere. Unceasing braiding of the magnetic field by convective motions yields an upward energy flux that eventually maintains a hot corona. We calculate the horizontally averaged density, which provides a density profile as a function of height that is more consistent with the density stratification of the real solar atmosphere.

The mesh of a zero-$\beta$ model has 256 grid points in the $x$ and $y$ directions, with a horizontal grid spacing of 576 km. This grid spacing is sufficient to capture the relevant magnetic structures in this study, for example, the magnetic connections between the sunspots. The same setup has also been used to model magnetic flux rope in the flare-productive region AR11158, as shown in the accompanying paper of this series \citep{Chen:2026}. The simulation domain covers an area of 147.456$^{2}$\,Mm$^{2}$, whereas the original observation covers an area of 276.480$^{2}$\,Mm$^{2}$. Thus, the horizontal domain represents a region that is shrunken by a factor of $f_{\rm V}$=1.875 from the actual size of AR11640. 

Such a spatial downscaling factor allows the model to cover the entire active region (with padding in the periphery), while still retaining a reasonable horizontal resolution and number of grid points. Similar approaches have been employed in previous works: e.g., a factor of 3.75 in \citet{Gudiksen+Nordlund:2005a} and 5.0 in \citet{Bingert+Peter:2011}. The evolution of volume-integrated magnetic energy in downscaled and original-scale models has been tested in \citet[][see Appendix therein]{Chen:2026}. The energy in the downscaled model multiplied by $f_{\rm V}^3$ matches very closely to that in the original-scale model, which means that the downscaled model represents the same energy density. Therefore, we expect the downscaled model in this study to provides a similar volumetric heating rate, although the exact impact of the spatial scaling needs to be investigated by a one-to-one scale model as done in \citet{Chen:2026}.

Given the same volumetric heating rate, downscaling the domain reduces the height of coronal loops and thus facilitates mass filling of loops. This implies that in the one-to-one scale model, heating long and high coronal loops appears even more challenging than it already is in the current setup. Conversely, the classical scaling law \citep{RTV} indicates that loop density and apex temperature, and hence the observables, depend not only on the volumetric heating rate but also on the loop length. Therefore, under a comparable volumetric heating rate and profile along loop, longer loops are predicted to reach slightly larger apex temperatures and densities. This might help explain the current differences between models and observations shown in \sectref{sec:result}. Although a quick estimate of the hydrodynamic properties could be obtained with hydrodynamic loop models \citep[e.g.,][]{EBTEL}, we expect to quantify these changes more accurately in a proper one-to-one scale model.

In the vertical direction, the domain is resolved by 1152 grid points and a grid spacing of 64\,km, reaching a height of 73.728\,Mm. The grid spacing follows the standard option for MURaM simulations with a corona and is used in the radiative MHD simulations described in \sectref{sec:rmhd}. This option achieves a reasonable balance between resolving the atmospheric stratification and affordable computational expenses.

The horizontal boundaries are periodic for all quantities. At the top boundary, the horizontal velocities are symmetric, and the boundary is open (symmetric) for upflows and closed (anti-symmetric) for downflows. The magnetic field matches a potential at the top boundary. The bottom boundary is set as the photosphere of the model, where the magnetic field follows the driving of the imposed electric field. The bottom boundary is closed for mass flows and symmetric for horizontal velocities.

The calculation starts from 2012 December 30 at 12:00 UT (Day 0.5 hereafter), when a pair of sunspots has appeared in the photosphere. A potential field calculated from $B_{z}^{\rm obs}$ at the bottom boundary and the assumption of a vanishing field at infinity serves as the initial condition for the magnetic field. Moreover, all the velocity components are zero. The start time remains in the very early stage of active region emergence and  skips those data when it is closer to the solar limb. The model is evolved for more than 3.5 solar days and terminated shortly after 2013 January 3 at 00:00 UT.

The primary purpose of the zero-$\beta$ model is to construct an evolution of the coronal magnetic field, from which radiative MHD models that consider more sophisticated energy transport processes can initiate. We expect that the coronal magnetic field evolves at a much faster pace than the flux emerges in the photosphere. Thus, to make the model of a long temporal evolution less computationally demanding and more doable, we accelerate the temporal evolution of the bottom boundary driver by a factor of $f_{\rm sp}$. This means that the magnetic field at the bottom boundary changes at a cadence of $720/f_{\rm sp}$; thus, the electric field derived from this enhanced cadence is also larger by the same factor. The factor $f_{\rm sp}$ also applies to $\Omega$ such that the total amount of rotation added to the system remains the same as that without accelerated evolution. The consistency of runs with different speed up factors has been tested and presented in \citet[][Appendix therein]{Chen:2026}. In the zero-$\beta$ simulations shown in this study, we use $f_{\rm sp}=12$ (i.e., an update cadence of 60\,s), which allows the model to cover the original 3.5-day evolution with approximately $7\times10^{5}$ time steps. We note that the speed up factor is {\emph not} used for the radiative MHD models described below. 

The vertical magnetic field in the photosphere (first layer in the computational domain) of the zero-$\beta$ model and observation ground truth (already rescaled) are displayed in \figref{fig:boundary}. The values along the blue dashed lines in the magnetograms are compared in the bottom panels. The zero-$\beta$ model is able to capture the evolution of the sunspots. Regions with sharp/small magnetic features in the observation appear more diffusive in the data-driven model, primarily because of the compromised horizontal resolution. The 2 snapshots correspond to approximately the middle and end of the evolution of the zero-$\beta$ model, indicating a consistent performance of the data-driven method through time. The zero-$\beta$ model shown here is calculated with $\Omega=-3\times10^{-6}$\,s$^{-1}$, which demonstrates that, as expected, the vertical magnetic field is not affected by the noninductive component.

\subsection{Radiative MHD Model}\label{sec:rmhd}
The radiative MHD models are the primary production runs in this study. We solve fully compressible MHD equations with a coronal energy balance as described in \citet{Chen+al:2023}. This set of equations considers realistic energy transport processes such as field-aligned thermal conduction and optically thin radiative loss and heating of the plasma by dissipation of kinetic and magnetic energy. Therefore, these simulations can provide realistic and detailed plasma properties in the model corona that can be used to inspect the plasma in the real active region via quantitative comparisons between model-synthesized observable and real remote-sensing observations.

With the hybrid model strategy, we can conduct radiative MHD simulations that are initialized at any time of interest. In this study, we run radiative MHD models that start from 2012 December 31 at 00:00 UT (Day 1, hereafter), 2013 January 1 (Day 2, hereafter), 2013 January 2 (Day 3, hereafter), and 2013 January 3 (Day 4, hereafter), respectively.

The simulation domain of a radiative MHD run has the same $147.456^{2}\times73.728$\,Mm$^{3}$ size as the corresponding zero-$\beta$ model. In the vertical direction, the 1152 grid points and the spacing of 64\,km remain the same. The refinement of the mesh is performed in the $x$ and $y$ directions, which are now resolved by 512 grid points and have a finer 288\,km grid spacing.

The boundary conditions are similar to those used in the zero-$\beta$ model, except for the mass flux at the top boundary. The mass flux in the uppermost 2 cells in the computational domain is copied symmetrically to the ghost cells and multiplied by a factor of 0.5 (first ghost cell) and 0.25 (second ghost cell), which is identical to the setup used in \citet{Chen+al:2022}. The thermal conduction flux at the top boundary vanishes. We note again that in the radiative MHD model, no $f_{\rm sp}$ is applied. The magnetic field at the bottom boundary is driven by the imposed time-dependent boundary electric field that follows the original 720\,s cadence of the observation data.

The initial condition of a radiative MHD run is set as follows. We adapt 3D cubes of all MHD variables from the snapshot of the quiet Sun simulation mentioned above (i.e., the snapshot that provides the density profile for zero-$\beta$ models). These cubes (1152 grid points starting from the mean optical depth unity layer) are used as the initial values for the density, energy, and velocity vectors. 

The quiet Sun magnetic field ($\mathbf{B}_{\rm QS}$) is added to the magnetic field of the zero-$\beta$ snapshot ($\mathbf{B}_{\rm z\beta}$) via the following steps to construct the initial condition for the magnetic field ($\mathbf{B}_{\rm init}$). We first calculate the potential field $\mathbf{B}_{\rm p-256}$ from the vertical component of $B_{\rm z\beta}(z=0)$, with a periodic horizontal boundary and vanishing field at infinity. It is straightforward to obtain the nonpotential components of the magnetic field in the $256\times256\times1152$ mesh of the zero-$\beta$ model by 
$$
\mathbf{B}_{\rm NP-256}  = \mathbf{B}_{\rm z\beta} - \mathbf{B}_{\rm P-256}.
$$ 
Then $\mathbf{B}_{\rm NP-256}$ is interpolated by the nearest neighbor method in the horizontal direction to a $512\times512\times1152$ mesh to fit with that of the radiative MHD model. We also use $B_{z}^{\rm obs}$ (the $512\times512$ resolution dataset) at the time when the radiation MHD model is initiated to calculate the high-resolution potential field ($\mathbf{B}_{\rm P-512}$). Finally, the initial condition of the magnetic field is given by 
\begin{equation}
\mathbf{B}_{\rm init} = \mathbf{B}_{\rm QS}+\mathbf{B}_{\rm P-512}+\mathbf{B}_{\rm NP-512}.
\end{equation}
Furthermore, a factor given by 
\begin{equation}
f_{v} = \frac{{\mathbf{B}_{\rm crit}^2}}{\mathbf{B_{\rm init}}^2+\mathbf{B}_{\rm crit}^{2}}
\end{equation}
is applied to the initial velocity field. In this study, we use $|\mathbf{B}_{\rm crit}| = 1500$\,G. This factor approaches unity in the weak magnetic field region and strongly reduces the velocity in the umbra areas of the sunspots and the atmosphere above. 

The radiative MHD models evolve the QS atmosphere with rather stochastic small-scale features to a loop-dominated active region corona. The transition time can be estimated by the time scale of evaporation flows filling the longest loops in the domain, which is approximately 2000\,s.\footnote{The height of the domain $L_z=73.728$\,Mm leading to the longest loop length of approximately $\pi L_{z}$. Given a typical coronal plasma velocity of 100\,km/s, we can obtain a loop filling time scale of 2300\,s.} This estimate is consistent with the results given by inspecting the mean coronal density and temperature as a function of time. Therefore, in the analysis and results shown in this paper, we exclude the first hour after the radiative MHD model is initiated.

\subsection{Control Experiments}\label{sec:control}
In addition to the run cases with 3 options of $\Omega$ described earlier in this paper, we also conduct more runs to investigate the impacts of the numerical resolution and initial conditions of the radiative MHD models on the active region corona formed. These control experiments are not sufficient to provide comprehensive coverage of the parameter space but do help evaluate the robustness of the strategy described above and its potential in routine applications to simulate real active regions.

\subsubsection{High resolution runs}
We perform high resolution runs for the radiative MHD models that start from Day 2 of the zero-$\beta$ runs with $\Omega = 0$ and $-3\times10^{-6}$\,s$^{-1}$. The high resolution runs are initiated by interpolating the $512\times512\times1152$ cubes of the corresponding radiative MHD runs that have evolved from their initial condition for $10^{5}$ iterations ($\approx$69\,min) to $1024\times1024\times1152$ by the nearest neighbor method. Therefore, the mesh has grid spacings of 144\,km and 64\,km in the horizontal and vertical directions, respectively. A horizontal electric field fitting the refined horizontal mesh is calculated as described above to drive the high resolution runs. The start time skips the initial relaxation stage of the standard resolution models, in which the quiet Sun corona evolves to the active region corona, to save computational resources. Changing the resolution of a snapshot also requires some relaxation time, which is much shorter. Both high resolution runs are evolved for approximately $2\times10^{5}$ iterations ($\approx$140\,min), and we use the data after $5\times10^{4}$ iterations for the analysis of the high resolution runs. 

\subsubsection{Uniform plasma initial condition} 
The quiet Sun that is used as the initial condition is filled with small-scale density and temperature structures and has a fully developed turbulent velocity field. These factors certainly give rise to extra structures and energy input in the active region corona; for example, when the quiet Sun velocity field is imposed on the active region magnetic field, although we do not expect this to be a major source of energy input in the radiative MHD models.

This control run is initiated from Day 2 of the zero-$\beta$ run with $\Omega=-3\times10^{-6}$\,s$^{-1}$. The horizontally averaged plasma properties of the quiet Sun snapshot (i.e., a horizontally uniform stratification) are used as the initial condition, with a vanishing velocity field. The other conditions are the same as those of a standard radiative MHD run. Thus, a quiet Sun magnetic field is still applied in the initial magnetic field; otherwise, virtually no energy input is given to support the padded area outside the observed active region.

\subsubsection{Upflow top boundary} 
The top boundary allows both outflow and inflows with a strong damping factor. This follows the setup used in previous MURaM simulations with a noneruptive corona \citep{Rempel:2017,Chen+al:2022}. In simulations where an eruption is expected \citep{Chen+al:2023,Chen:2026}, the top boundary is open (symmetric) for outflows and closed (anti-symmetric) for inflows to facilitate outward propagation of, e.g., an erupted magnetic flux rope. We refer to the latter "upflow top boundary" hereafter.

This control run is initiated from Day 2 with $\Omega=-3\times10^{-6}$\,s$^{-1}$ and employs all parameters and setups that are identical to those of the standard radiative MHD run, except for an upflow top boundary.

\subsection{Summary of Cases}
To summarize, all simulation runs presented in this paper are listed in \tabref{tab:list}. In the table, "Bevo" refers to zero-$\beta$ runs that evolve the coronal magnetic field over the course of active region emergence. Runs named "D" are radiative MHD models that are started on a certain day, as previously defined. The $\Omega$ values indicates from which zero-$\beta$ model (as well as the driver electric field) they are constructed. Suffixes "High", "Uniform", and "Upflow" stand for the control experiments with high resolution, with horizontally uniform initial density and temperature, and with an upflow top boundary, respectively.

\section{Results}\label{sec:result}
\subsection{Emergence of the Active Region}

\begin{figure*}[ht!]
\center
\includegraphics[width=15.3cm]{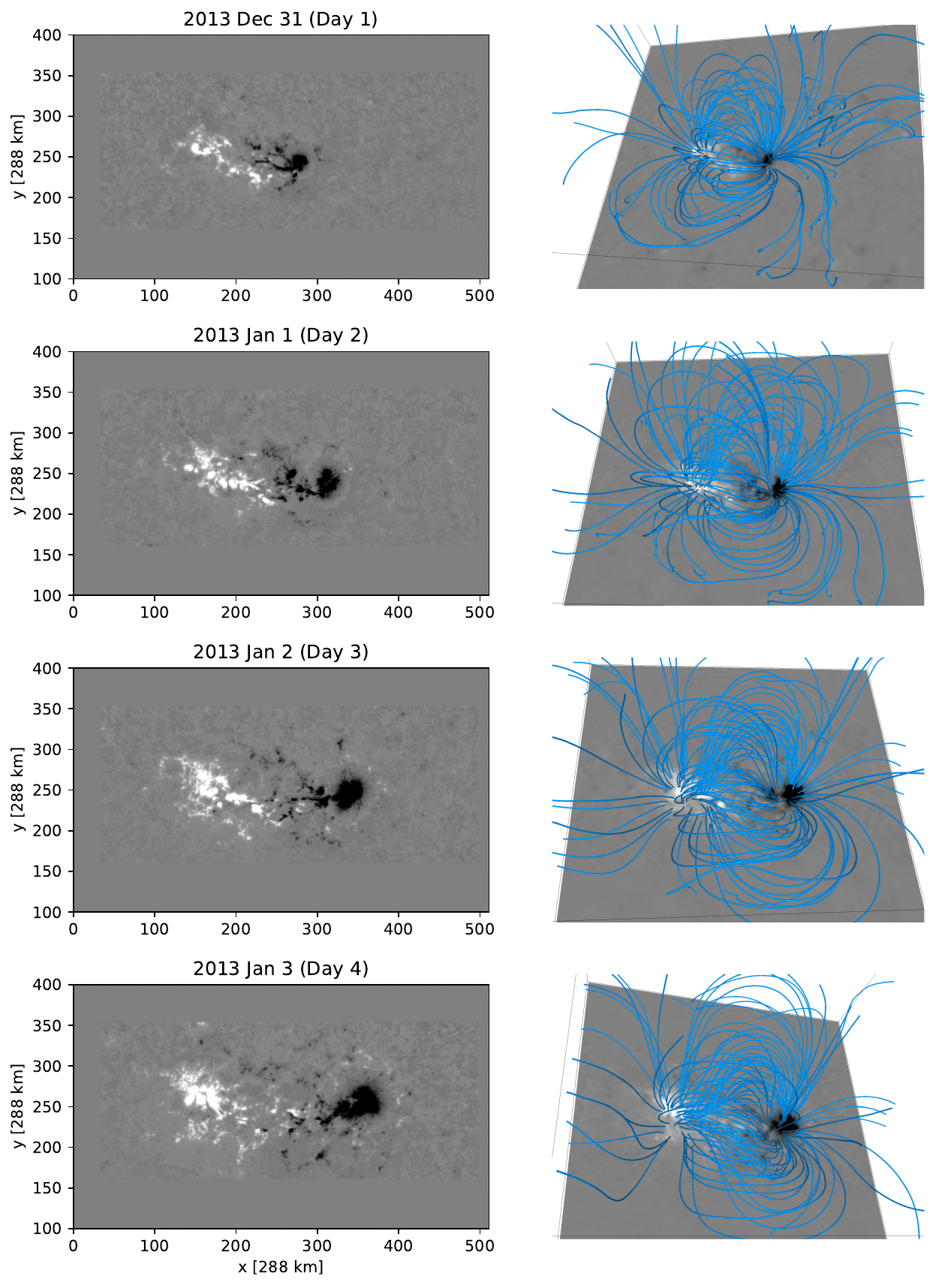}
\caption{The left column shows the evolution of the observed radial magnetic field of AR11640. Only the central part of the padded array is displayed. The right column presents the coronal magnetic field in the Bevo\_$\Omega$0 run. The angle of view in each panel is set to reflect the position of the active region on the solar disk at the observed time. The grayscale images show $B_{z}$ at the bottom of the simulation domain. Magnetic field lines are calculated from static seed points that are uniformly distributed in the central part of the domain.
\clearpage
\label{fig:mag}}
\end{figure*}

The target active region in this study, AR11640, is a calm active region that has a relatively simple bipolar magnetic configuration. It is not flare-productive, although some microflares occurred. The main emergence stage of this active region is observed across the solar disk, which is why it serves as a usable test case of the method described here. 

The evolution of the observed vertical magnetic field is shown in \figref{fig:mag}. The two sunspots demonstrate separation instead of rotation or shear motions, which are known to result in rapid increases in the free magnetic energy that causes major eruptions.

The evolution of the corona magnetic field lines in the lowest energy case ($\Omega0$ model) in the zero-$\beta$ runs is shown in the right column of \figref{fig:mag}. The seed points are identical in all 4 panels and are uniformly distributed in the central part of the domain covering the two sunspots and in the height range up to 40\,Mm. As the active region emerges, more magnetic loops connecting the two sunspots appear, particularly in the higher part of the domain. Long but low-lying loops are also formed beginning on Day 2. We note that field lines may connect across the periodic horizontal boundary.

Comparisons between field lines in a numerical simulation and observed EUV loops have often been performed in previous investigations. However, the magnetic field permeates in space, and in contrast, EUV loops are discrete individual structures that indicate inhomogeneity in the coronal plasma and, eventually, heating in space. Therefore, a comparison between the simulated active region and observations needs to be performed with radiative MHD models that yield appropriate observables.

\subsection{Comparison of the observed and model coronae}

\begin{figure*}[ht!]
\includegraphics[width=18cm]{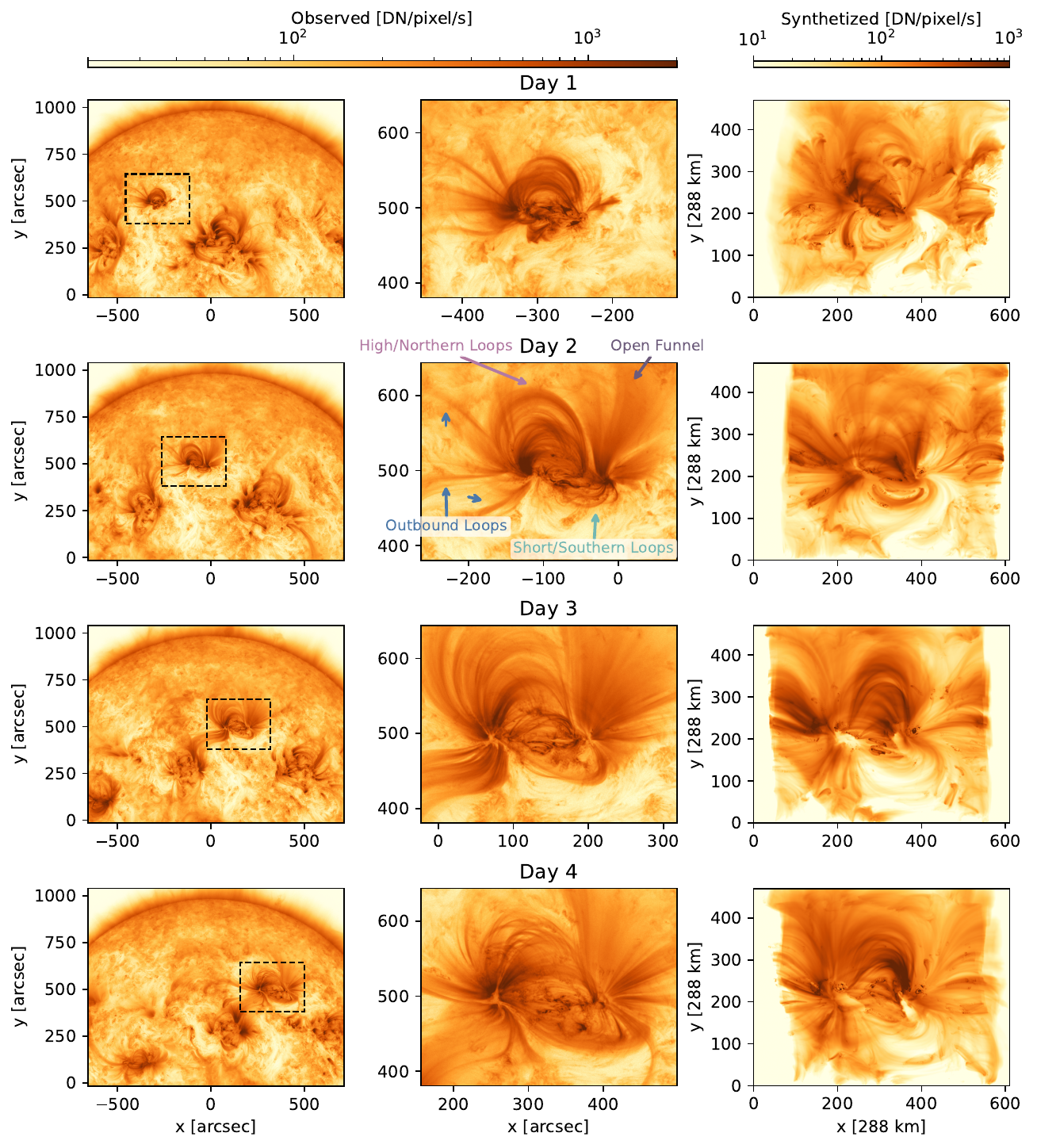}
\caption{Comparison of the observed (left and middle columns) and synthesized (right column) AIA 171 \AA~images of AR11640  over 4 days (each row). The actual AIA 171 \AA~images captured at 2:00 UT are displayed on a logarithmic scale between 20 and 2000 DN/pixel/s. The black boxes in the left column indicate the field of view of the images in the middle column. The colored arrows indicate key coronal features for comparison in the rest of the study. The synthesized 171 \AA~images from the radiative MHD models with $\Omega=0$ (see main text for details) are shown on a logarithmic scale between 10 and 1000 DN/pixel/s, which is lower than that of the observed images. The angles of view of the synthesized images are chosen according to the locations of the actual active region on the solar disk on the corresponding days. 
\label{fig:aia}}
\end{figure*}

\begin{figure*}[ht!]
\center
\includegraphics[width=16.1cm]{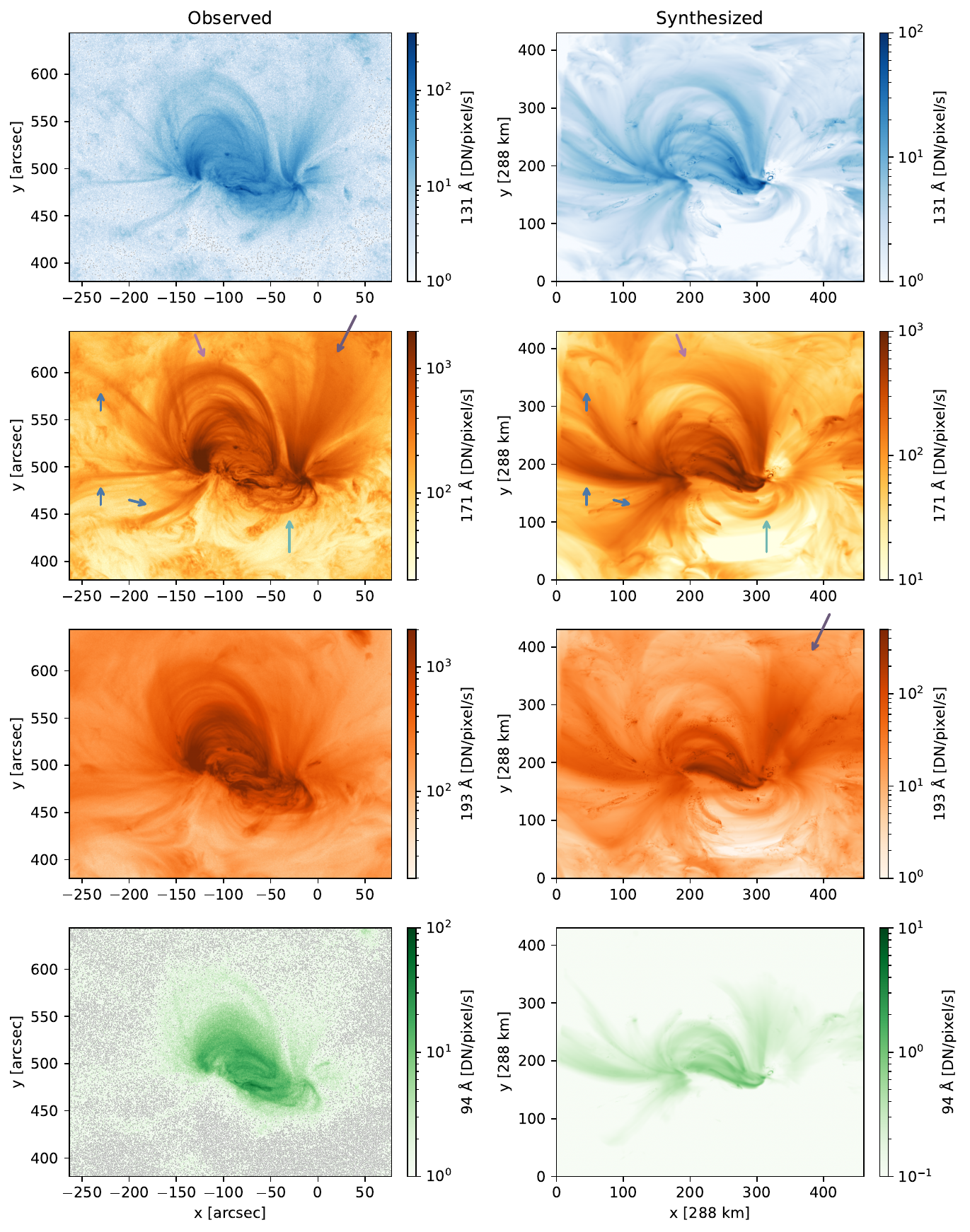}
\caption{Comparison of the observed and synthesized AIA images in 4 passbands. The actual AIA 131 \AA, 171 \AA, 193 \AA, and 94 \AA~images captured on Day 2 at 02:00 UT are displayed in the left column. The synthesized images of the corresponding channels from the D2\_$\Omega$3 radiative MHD models are shown in the right column. The angle of view of the synthesized images is chosen according to the locations of the actual active region on the solar disk on Day 2. The arrows are identical to those shown in \figref{fig:aia} and indicate key features for comparison. Note that the dynamic ranges of the images are chosen to achieve a visual similarity. A quantitative comparison of the count rates is presented in \figref{fig:countrate}.
\label{fig:aia_d3}}
\end{figure*}

\begin{figure*}[ht!]
\includegraphics[width=18cm]{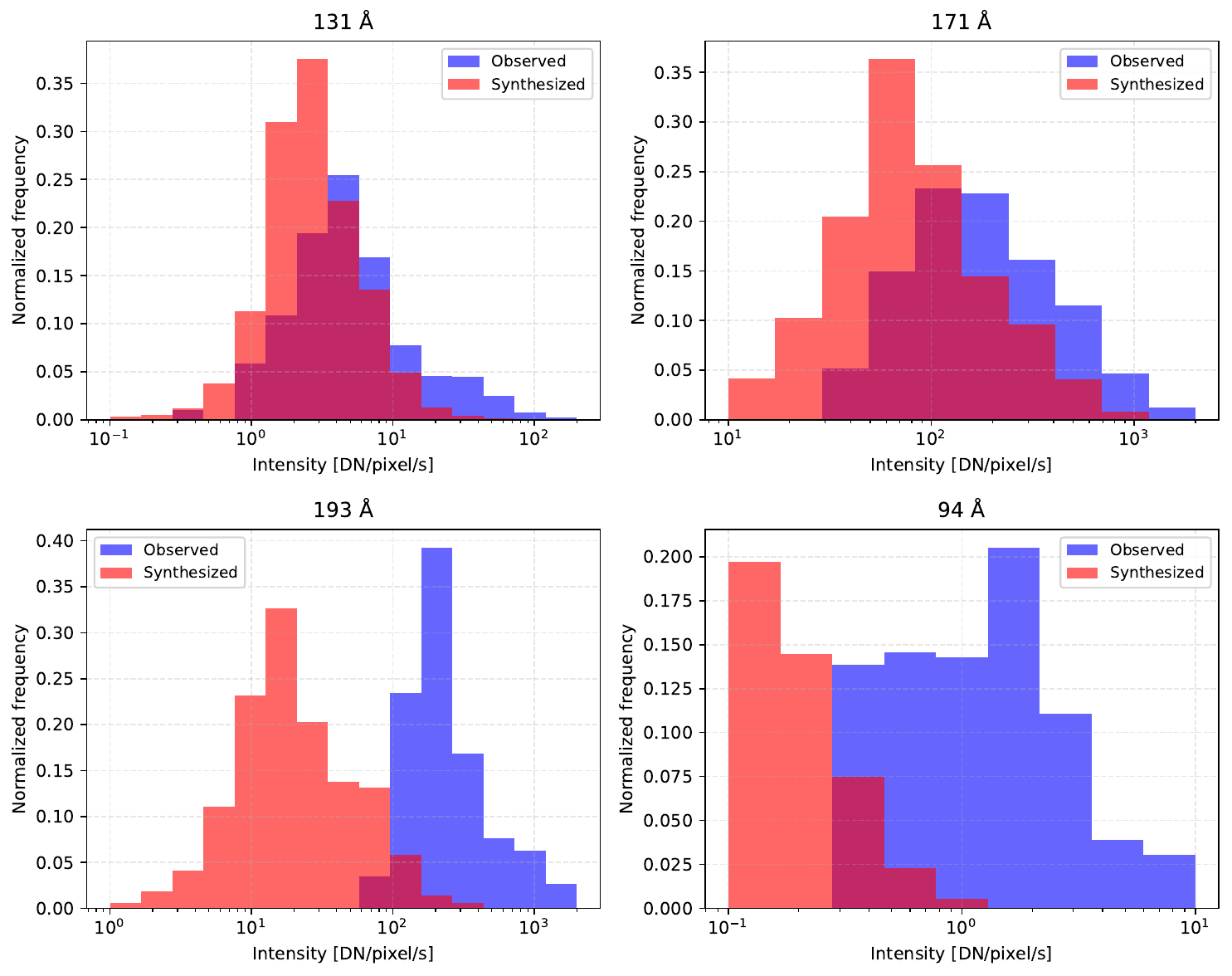}
\caption{Histograms of the observed and synthesized intensities in 4 AIA channels, which are shown in \figref{fig:aia_d3}. For each channel, the histogram is calculated for the logarithmic intensity and normalized to the numbers of pixels/grid-points in the observed/synthesized images. Note that the model-synthesized intensity may contain arbitrarily small numbers. A lower bound is applied to each channel as indicated by the axis.
\label{fig:countrate}}
\end{figure*}

We compare the coronal EUV emission synthesized from the density and temperature in the radiative MHD models with actual observations of AR11640. This analysis assesses the extent to which the models may resemble observations and sheds light on the plasma properties that are difficult to probe in remote sensing observations.

\subsubsection{The observed and model coronae over 4 days}
The AIA 171 \AA~channel images of AR11640 at 02:00 UT of each day are displayed in the left and middle columns of \figref{fig:aia}. The level-1 AIA images have been corrected for the time-dependent degradation and exposure time, which is applicable for images of other channels shown later in this paper. For comparison, the right column shows synthesized AIA 171 \AA~images from the radiative MHD models of $\Omega 0$ on these 4 days. The synthesized images used a temperature response function generated by the {\it aia\_get\_response} procedure in the {\it SolarSoftware} package with ({\it /DN, /TEMP}) parameters and a photospheric abundance, which is consistent with that used for the optically thin radiative loss in the simulation.

As described in \sectref{sec:method}, these models are not from one single simulation that is evolved for 4 days but from 4 individual simulations, which are started at time instances that are 1 day apart in the $\Omega0$ zero-$\beta$ model. The snapshots are captured when the models are evolved for $2.5\times10^{5}$ iterations (more than 2 hours) from the initial condition. Both the real active region and the models show short time-scale variations in loop brightness and intermittent small-scale activities inside the active region as well as in its periphery. Here, the time stamp is arbitrarily chosen and does not aim to match the small-scale and transient variations. The angles of view of the synthesized images are set according to the positions of the real active region on the solar disk on these 4 days. Instead of normalized intensity, which limits quantitative comparison, the observed and synthesized AIA images are displayed in their original units.

On Day 1, when it is still in the early stage of active region emergence, the sunspots are closer to each other, and the coronal EUV features are mostly short closed loops in the active region core. The simulated active region also shows short loops connecting the two sunspots, although the detailed structures do not seem to match the observed loops in a one-to-one fashion. The observation indicates that several distinguishable outbound loops connect  from the trailing sunspot to the quiet Sun. A similar trend is also found in the model; however, the contrast of these loops to the background and the separation between individual loops are not as clear as those in the observation. The model uses a periodic horizontal boundary and imposes an extra small-scale magnetic field from the quiet Sun simulation that is theoretically generated and hence independent of this particular active region. These treatments may very likely change the connection of the field lines from the sunspot to areas outside the active region or computational domain.

On Day 2, in addition to the existing short loops between the sunspots, higher coronal loops are formed, as marked in \figref{fig:aia} as "High/Northern Loops".  A prominent open funnel, which may also be the leg part of large-scale closed coronal magnetic field lines, can also be seen above the leading sunspot. This feature is indicated by the arrow in \figref{fig:aia} with the annotation "Open Funnel". A group of short and low-lying loops connect from the leading sunspot to the center of the active region, which is also marked as "Short/Southern Loops" in \figref{fig:aia}. In the simulated active region, bright short loops remain clearly visible. Moreover, longer and higher closed loops are discernible but not as bright as those in the real active region. The model corona does not show a spread open funnel, although some thinner bright features can be seen above the leading sunspot. A group of low-lying short loops are also found in the model at the correct location (connecting the leading spot and active region center through the south) and in a similar shape (a few thin curved threads fan out from the sunspot) as in the observation. The outreaching loops from the trailing sunspot, as marked by the arrows in \figref{fig:aia} with the annotation "Outbound Loops", become clearer in the model on Day 2.

On Day 3 and Day 4, the active region is more developed and exhibits similar coronal structures. More long and closed loops are formed between the two sunspots. The open funnel evolves to large fan loops connecting the leading sunspot to regions far beyond the domain of the simulation in this study. The outbound loops also connect to another active region southeast of AR11640, which is clearly shown in the large field of view images in the left column. In general, reproducing these large and out-of-the-domain loops in the radiative MHD model is very challenging. This task requires a large and high domain that encompasses related active regions while still persevering a reasonable grid spacing to resolve small-scale magnetic flux concentrations in the quiet Sun and thus seems to be impractical. 

Nevertheless, $\Omega0$ radiative MHD models for Day 3 and Day 4 are performed. The synthesized AIA 171 \AA~images in \figref{fig:aia} show mostly bright loops in the center part of the active region. Because the sunspots are more separated in the later days, these loops are also higher than those seen on Day 1 and Day 2. Some long but low-lying loops in the southern half of the active region, which probably originate from the short/southern loops in the same area in the first 2 days, are reproduced in the models, albeit with a lower intensity. 

The long/northern loops are largely missing in the models. The cause can be illustrated by comparing the destinations of the loops starting from the trailing sunspot. In the real active region, these loops connect to the leading sunspot, forming large arches. However, loops that originate from the same place in the models tend to connect through the periodic horizontal boundary to the other side of the leading sunspot. This also explains the absence of a large open funnel above the leading sunspot, because the field lines are connected to the trailing sunspot through the boundary and form closed loops. 

We note that on Day 4, the similarity between the model and the observation is slightly improved, in the sense that the loops reaching the boundary of the model resemble the outbound loops and open funnel in the observations. This is still because the sunspots in the model are connected through the periodic horizontal boundary, which coincidently mimics the effect of inter region connections in the observations.

\subsubsection{The observed and model coronae across 4 passbands}
A more critical assessment of the consistency between observations and models is the emission from plasma across a wide temperature range. From the $\Omega3$ model on Day 2, we calculate the emission for the AIA 131 \AA, 171 \AA, 193 \AA, and 94 \AA~channels, and the synthesized images are compared with the observations in \figref{fig:aia_d3}. The 131 \AA~channel prefers plasmas that are cooler than the peak temperature of the 171 \AA~channel. Compared with the 171 \AA~channel, the 193 \AA~channel has a higher peak response temperature and displays emission from hotter plasmas. The 94 \AA~channel captures emission from super hot plasma above 6\,MK and for noneruptive active regions, such as the one studied here, is mostly detected in the active region core. 

Key EUV features observed in the real active region, which are marked by arrows that are identical to those in \figref{fig:aia}, can also be found in the corresponding synthesized images. Particularly in the channels for lower temperatures, the model reproduced the long/northern loops, as well as the short/south loops between the sunspots and the outbound loops from the trailing sunspot. The prominent open funnel observed in the 131 \AA~and 171 \AA~channels is better seen in the 193 \AA~channel of the synthesized images, as indicated by the arrow. The 193 \AA~channel reveals similar diffusive emission, whereas the brightness of the long/northern loops becomes much lower in the model. Last but not the least, reproducing the observed 94 \AA~channel emission is the most challenging. The synthesized 94 \AA~emission appears to have a similar shape to that of the lower temperature channels, indicating that the model does not have enough hot plasma and that the signal only comes from the cool tail of the response function.

It is necessary to note the dynamic ranges of the synthesized images in \figref{fig:aia_d3} are tuned to achieve the best visual comparison. The intensity of the synthesized images is on average lower than that of the observed images by at least a factor of 2 (e.g., 171 \AA~channel). A quantitative comparison between the observed and synthesized intensities in the 4 AIA channels is presented in \figref{fig:countrate}. The histograms of the intensity are computed with an interval of 0.2 on a logarithmic scale, and the frequencies are normalized by the total number of pixels, respectively.

The statistics appear consistent with the visual comparison in \figref{fig:aia_d3}. The distributions of the synthesized 131 \AA~and 171 \AA~intensities reproduce a consistent shape as the observation ground truth does, with a shift toward to a lower count rate. In contrast, the difference in the hotter channels is more obvious. The finding reveals that the model active region has significantly less hot plasma in the active region core. The insufficient emission from the diffusive corona, as shown in the 193 \AA~images, may be due to two aspects: the heating is not strong enough to fill long field lines that reach the higher part of the domain, and the limited domain size may reduce the accumulation along the line of sight.

\subsection{Coronal Plasma Properties in the Control Experiments}

\begin{figure*}[t!]
\includegraphics[width=18cm]{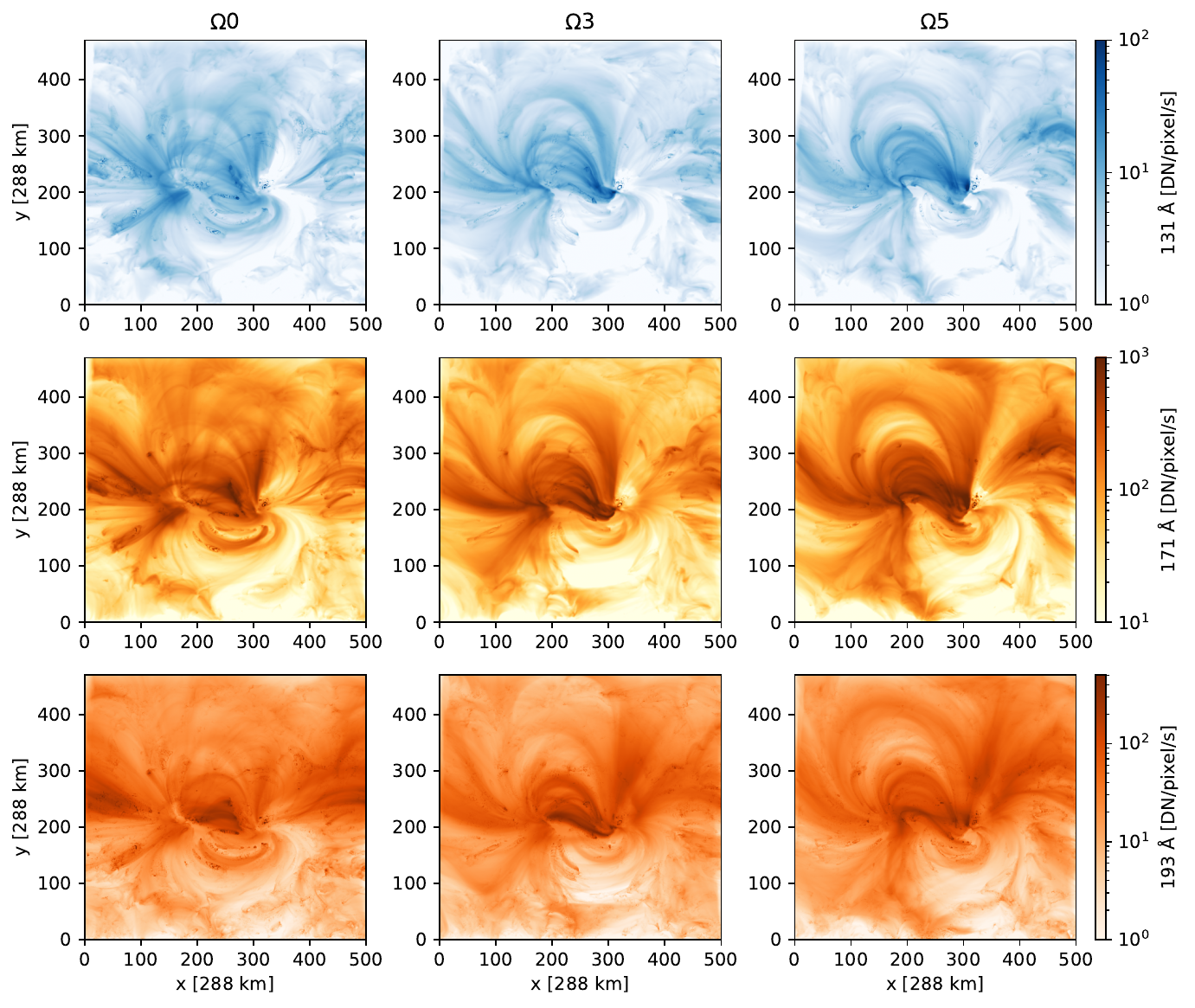}
\caption{A comparison of synthesized EUV emission from run cases on Day 2 with different $\Omega$ that adds an additional twist in the magnetic field while keeping the vertical component unchanged (see main text for details). Each column presents the results from a certain $\Omega$ value. The 3 rows show AIA 171 \AA, 131 \AA, and 193 \AA~images, respectively.
\label{fig:dive}}
\end{figure*}

\begin{figure*}
\includegraphics[width=18cm]{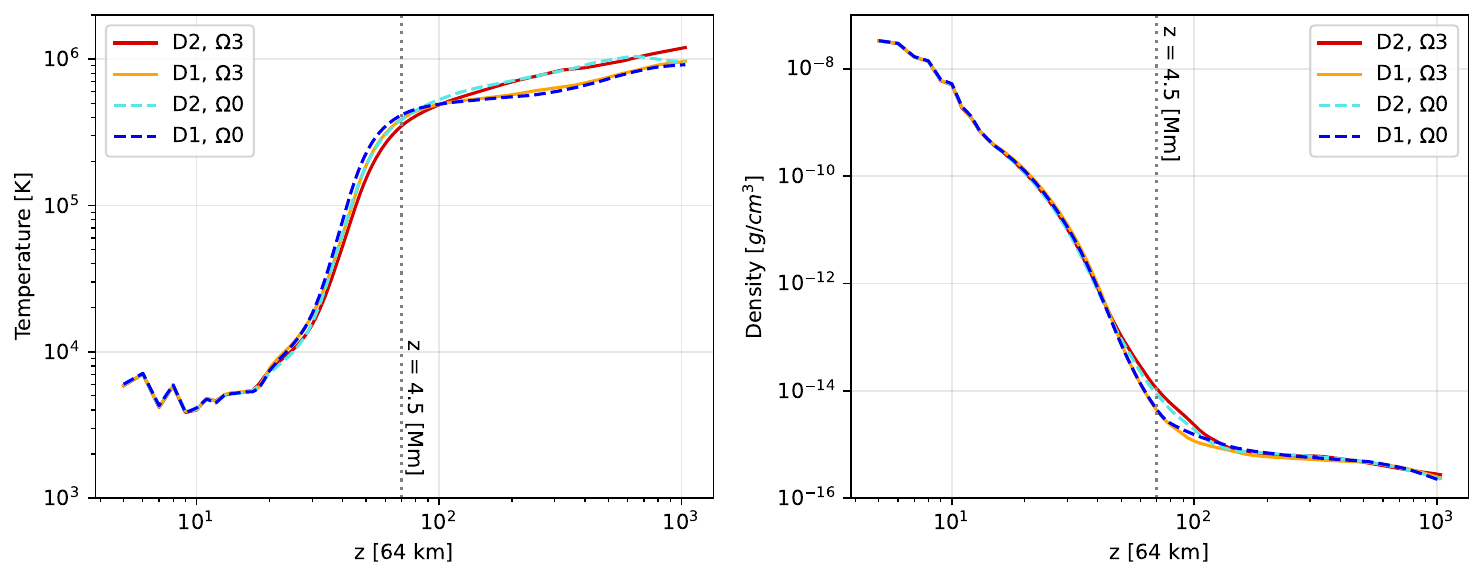}
\caption{Coronal density and temperature as a function of height. The results from 4 run cases, as indicated by the legend, are compared. The data are averaged over time for a period of more than 1 hour. The 3D cube is averaged in the horizontal dimensions, which provides the height profile shown here. The axis of height is displayed on a logarithmic scale, such that the lower atmosphere of a stronger stratification is stretched, whereas the coronal part with a much larger scale height is compressed. The vertical dashed line is placed at 70 grid points (4.48\,Mm) above the bottom boundary and indicates the bottom of the corona or say the top of the transition region.     
\label{fig:rhot_z}}
\end{figure*}

\begin{figure*}
\includegraphics[width=18cm]{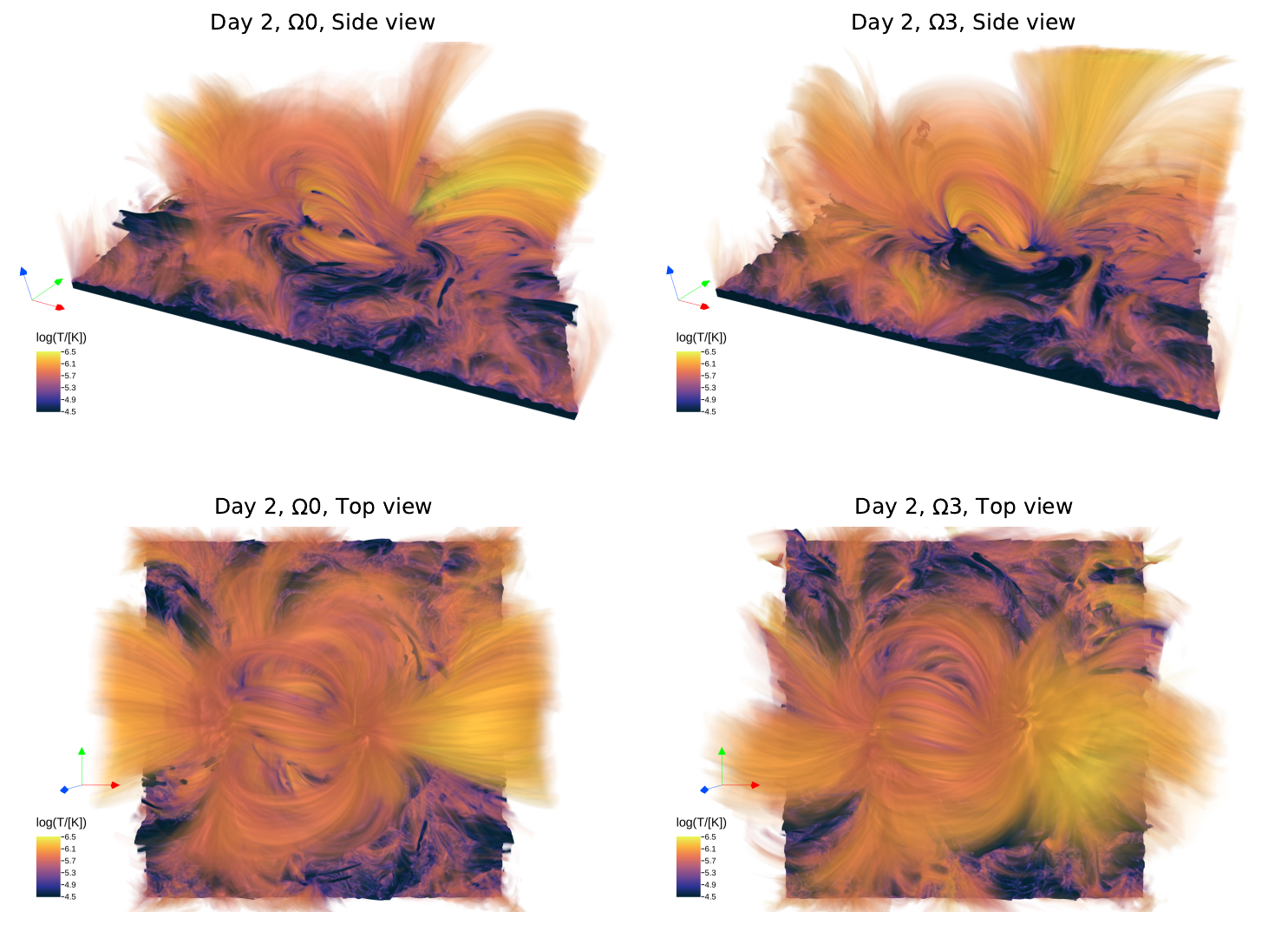}
\caption{A 3D rendering of the coronal density and temperatures in the $\Omega0$ and $\Omega3$ models on Day 2. The opaque features display the plasma density. Only the density values of the loops connecting the sunspots are illustrated, by forcing lower values in the coronal volume to be completely transparent. The density features are colored according to their temperature, as indicated by the color bar. The top and bottom rows show an inclined side view and a top-down view, respectively.
\label{fig:rhot}}
\end{figure*}

\begin{figure*}[t!]
\includegraphics[width=18cm]{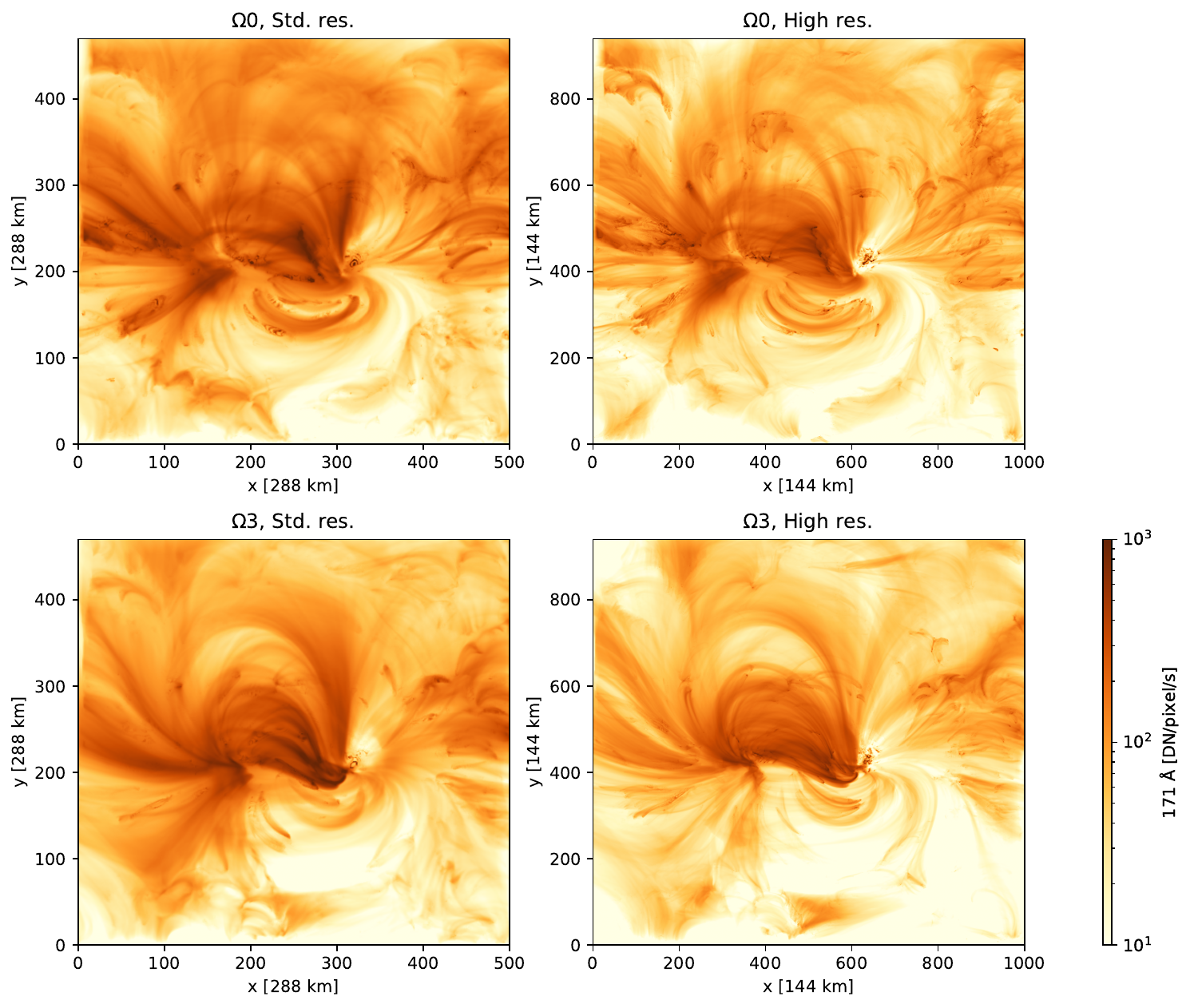}
\caption{Synthesized AIA 171 \AA~images of the radiative MHD models (on Day 2) with standard and high resolutions. The left column shows the $\Omega0$ and $\Omega3$ models calculated with standard horizontal resolution (512 grid points $\times$ 288\,km grid spacing). The right column displays the models calculated with the same parameters but a high resolution mesh (1024 grid points $\times$ 144\,km grid spacing). The vertical grid spacing for all the models is 64\,km.
\label{fig:hires}}
\end{figure*}

\subsubsection{Model corona with different nonpotential magnetic fields}
We investigate how the model corona changes when different nonpotential magnetic fields are imposed by a nonvanishing $\Omega$ parameter. $\Omega=0$ does not mean that the corona magnetic field is potential, as the evolution of the radial magnetic field involves motions that can inject free magnetic energy into the simulation volume. A nonvanishing $\Omega$ can be intuitively understood as applying a rotation to the vertical flux tube. Given the calm nature of this active region, we only used small $\Omega$ values, which seem to help to resemble the real active region. More extreme values have been tested (e.g., $\Omega=1.0\times10^{-5}$, which is not shown in this paper). They give rise to results that obviously deviate from the behaviors of the observed active region, for example, the formation of a sigmoid that eventually erupts.

Comparison of the EUV images calculated from the radiative MHD models with $\Omega=0$, $3\times10^{-6}$, and $5\times10^{-6}$\,s$^{-1}$ on Day 2 are presented in \figref{fig:dive}. The long/northern loops and short/southern loops, as marked in \figsref{fig:aia} and \ref{fig:aia_d3}, shift (in other words, lean or are distorted) to the trailing (East) and leading (West) directions, respectively, and this trend becomes more evident in models with larger amplitudes of $\Omega$. This geometry illustrates that the magnetic field starting or ending in the sunspots deviates more severely from the potential field because of the added twists. Similarly, the outbound loops discussed in earlier sections become more curved as $\Omega$ increases. Finally, the overall brightness of the active region remains consistent in models with different values of $\Omega$. This is primarily because the $\Omega$ values considered remain reasonably small. The intensities of some loop structures (e.g., the long/northern loops) clearly increase in large $\Omega$ models. This indicates a significant localized increase in the heating rate in some loops, particularly those connected to the strong magnetic flux concentrations; this effect is somewhat expected, as the rotation is directly applied to $B_{z}$, as shown in \equref{equ:dive}.  
 
Overall, the $\Omega3$ model on Day 2, which has been used to make the comparison shown in \figsref{fig:aia_d3} and \ref{fig:countrate}, provides the highest similarity with the observation on that day, which means that a twisted magnetic field, albeit weak, exists in the real sunspots.
 
\subsubsection{Thermodynamical properties of coronal plasma}
Measuring the temperature and density remains a main task and challenge in observations of the solar corona. Plasma properties can by be deduced from, for example, the spectral line ratio or DEM analysis of EUV images of different temperature responses, but both methods suffer from overlapping structures along the line-of-sight due to the optically thin nature of the coronal plasma. On the other hand, if a realistic radiative MHD model can quantitatively reproduce EUV observations of a real active region, such a model may serve as an alternative or complement to the inversion method and can help to investigate the plasma properties/dynamics in the 3D coronal volume, which is even beyond the capability of spectroscopic observations.

The plasma density and temperature cubes of the radiative MHD models in a 4200\,s period (between $1.5\times10^5$ and $2.5\times10^{5}$ iterations that are well beyond the initial relaxation) are averaged in time and in the horizontal dimension. The mean density and temperature profiles of the 4 run cases are plotted in \figref{fig:rhot_z}. The mean atmosphere is composed of a cool and dense lower atmosphere, a transition region, where the temperature/density steeply increases/decreases, and an extensive hot and tenuous corona. This structure is consistent with previous radiative MHD models of an active region corona and is a robust result of the energy balance between the optically thin radiative cooling, thermal conduction, and coronal heating.

The coronal temperatures in all 4 cases are very close, particularly for cases on the same day but with different $\Omega$. A small difference can be found between the models on Day 1 and Day 2, and the latter presents a slightly hotter corona. This comparison indicates that the mean coronal heating rate, which ultimately controls the mean coronal temperature and density, is only weakly affected by the twists added to the magnetic field. This can be expected because only weak rotation is applied in this study to fit the behavior of AR11640 and is consistent with the intuitive comparison shown in \figref{fig:dive}. Moreover, a hotter corona in the Day 2 model suggests that coronal heating clearly depends on the magnetic flux of the active region. Like in previous works, these models also rely on dissipating the energy flux generated by the braiding of magnetic field lines by photospheric motions. Therefore, a stronger magnetic field tends to give rise to higher energy input in the corona and stronger heating. Another effect is that on Day 2, more magnetic field lines reaching the higher domain are formed, which helps to channel the energy flux to the higher part of the domain and gives rise to the longer loops that start to appear on the Day 2 models in \figref{fig:dive}.

The plasma density and temperature structures in the coronal volume are displayed in \figref{fig:rhot}. Here, we focus on models on Day 2, when more loops are formed. The transparency of the 3D rendering is chosen to visualize the loops connecting the two sunspots and the dense plasma in the lower transition region. The color coding reflects their temperatures. The plasma of lower density (e.g., the tenuous diffusive plasma filling the upper corona) is made transparent, and the plasma below the transition region is opaque in this visualization. 

In the lower part of the volume, we find a corrugated chromosphere and transition region. The middle height of the domain is dominated by plasma loops from 1 to 3 MK. The density features in the 3D space illustrate many more thin threads than the synthesized EUV images, which suffer from strong light-of-sight integration. These fine structures are manifestations of the highly inhomogeneous heating in the 3D space, which is shown later in the paper.

The top view of the outbound loops connecting the sunspots to the boundary clearly illustrates the effect of the twists applied in the models: the loops in the $\Omega3$ model are curved. Another noticeable impact of the twisted magnetic field is that the $\Omega3$ model yields a wider and hotter open funnel above the leading sunspot, which is a prominent structure observed in the AIA 171 \AA~images of the real AR11640. Here, in the model, the plasma in the open funnel has a higher temperature than the peak response of the 171 \AA~channel. This explains why this structure is better seen in the synthesized AIA 193 \AA~image of the $\Omega$3 model, as shown in \figref{fig:dive}. 

Therefore, although the current setup may not be able to account for all situations in an active region, the 3D coronal density and temperature structures in the radiative MHD models could reproduce a realistic atmosphere stratification and yield some particular key structures that are consistent with observations, merely under the confinement of the bottom magnetic field $B_{z}$ and an educated guess of free parameters $\Omega$.

\subsubsection{Model corona in high definition}
Although the standard horizontal resolution of 288\,km is already finer than the length to which an AIA pixel corresponds to, the nature of a multipoint scheme determines that resolving a structure requires more grid points in a numerical simulation than in observations. Furthermore, the dissipation of kinetic and magnetic energy that heats the coronal plasma may also depend on the grid spacing: a finer grid spacing leads to a lower resistivity and viscosity in the code \citep{Rempel:2014}, whereas a larger gradient of velocity and magnetic field can be built. Therefore, it is interesting to investigate how the radiative MHD models of AR11640 may change if a higher resolution is used.

We rerun simulations for the $\Omega0$ and $\Omega3$ models on Day 2 with a two times better horizontal resolution (144\,km). A comparison of the standard and high resolution models is presented in \figref{fig:hires}. The overall intensities of the synthesized AIA 171 \AA~images of the high resolution models are slightly lower than those of the standard resolution models, which is true for both $\Omega$ values. This implies a weak decrease in the mean coronal heating rate as the grid spacing decreases. We note that in this comparison, the standard and high resolution models use a bottom boundary driver that is calculated from identical observation data (although they are interpolated to match the meshes of different simulation cases). In models that include self-consistent magnetoconvection in the photosphere and beneath, the energy flux given by the magnetic and velocity fields on smaller scales can provide an extra heating contribution, which could compensate for the decrease in the heating, as we see in this comparison. This is supported by an experiment with simulations that cover the range from the uppermost convection zone to the corona, as presented by \citet{Lu+al:2024L}. When the grid spacing is refined (192\,km to 96\,km), the mean coronal properties remain unchanged.

The major difference caused by the resolution change is that the smooth structures found in the standard run cases exhibit more isolated fine threads in the high resolution rerun. This helps to resolve more individual loops in a loop bundle. For example, the high/northern loops in the standard $\Omega3$ model appear to be a wide and diffusive bundle where some low-contrast individual threads may be discerned, whereas in its high resolution rerun, narrow and high contrast threads overlaid on the diffusive background bundles can be resolved. Similar behaviors are found in the short/southern loops and the outbound loops.

\subsection{Plasma Dynamics and Waves}\label{sec:doppler}

\begin{figure*}[t!]
\includegraphics[width=18cm]{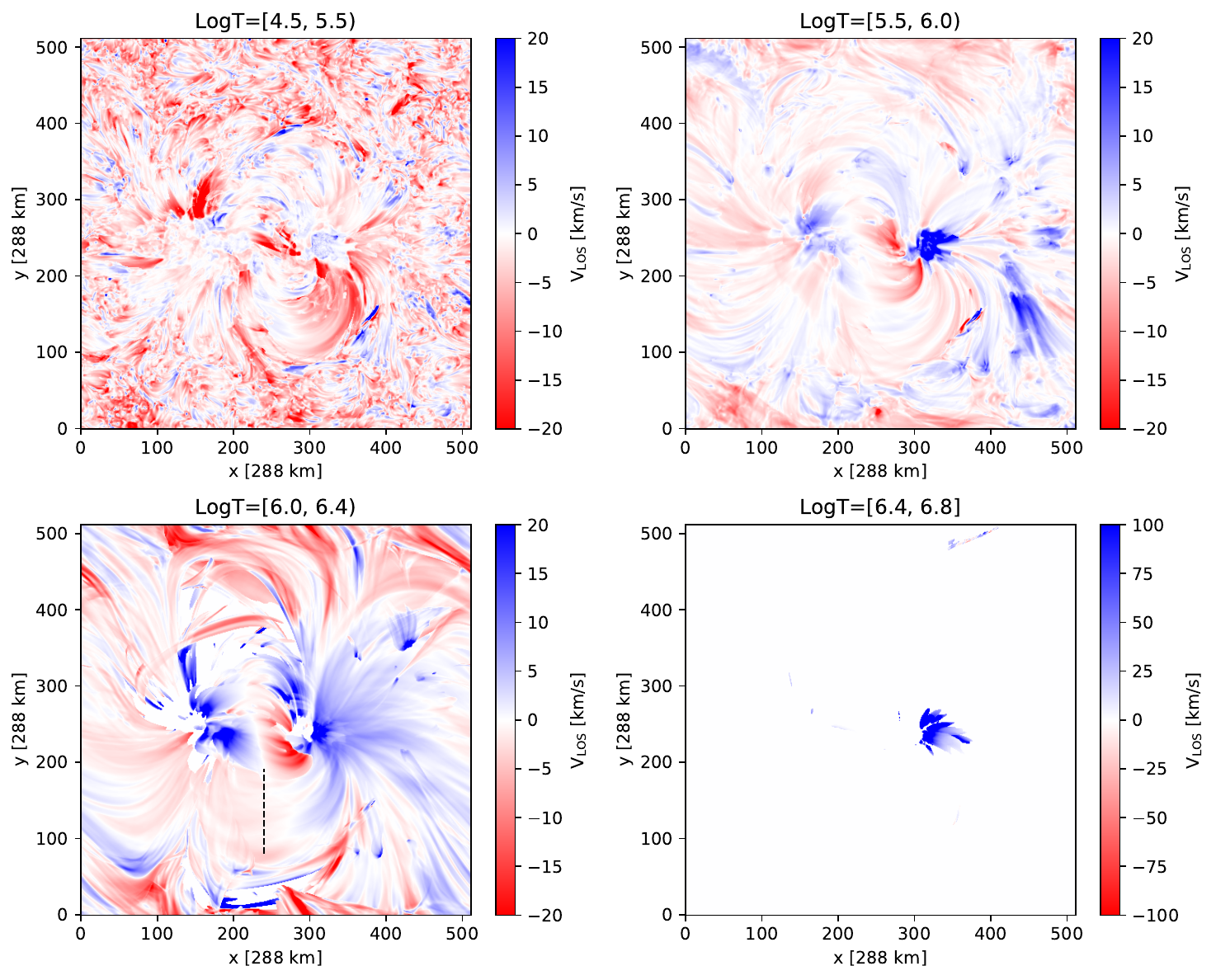}
\caption{Line-of-sight velocities of the plasma in different temperature ranges. The emission-measure-weighted mean velocity in the display temperature range is obtained following \equref{equ:emwv} and is equivalent to the Doppler velocity that would be measured from a spectral line forming in the given temperature range. Positive values (shown in blue) correspond to upflows. The vertical dashed line indicates the slit position for the time-distance diagram shown in \figref{fig:wave}.
\label{fig:doppler}}
\end{figure*}

\begin{figure*}[t!]
\includegraphics[width=18cm]{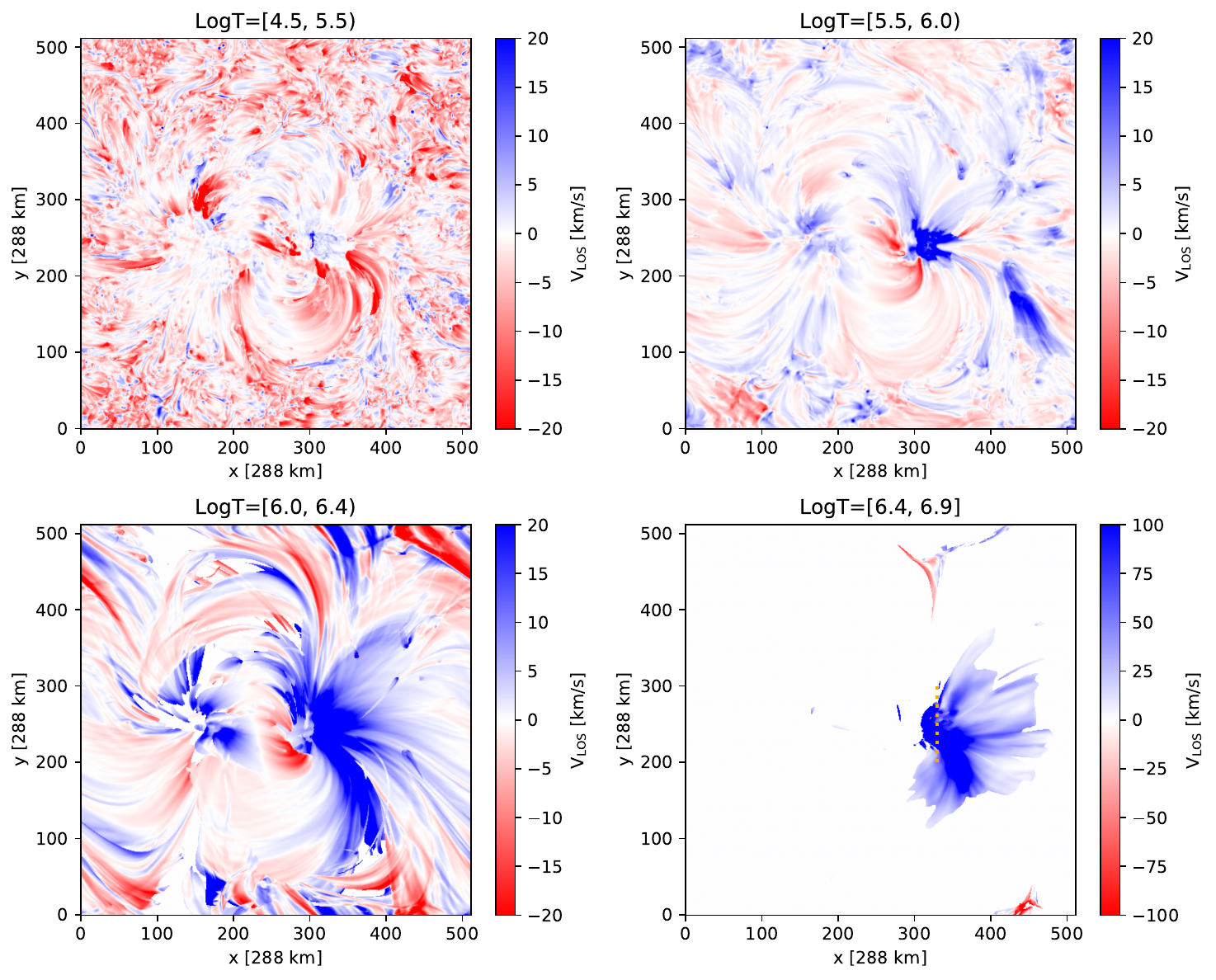}
\caption{Line-of-sight velocity for the model with an upflow top boundary (closed for inflows, see \sectref{sec:control} for details). The vertical dotted line indicates the slit position for the time-distance diagram shown in \figref{fig:wave_up}. An animation of this figure is available in the online version of the article. The animation runs for 20 s and covers a time period of 71 min starting from Day 2 at 00:35.
\label{fig:doppler_up}}
\end{figure*}

\begin{figure*}[t!]
\includegraphics[width=18cm]{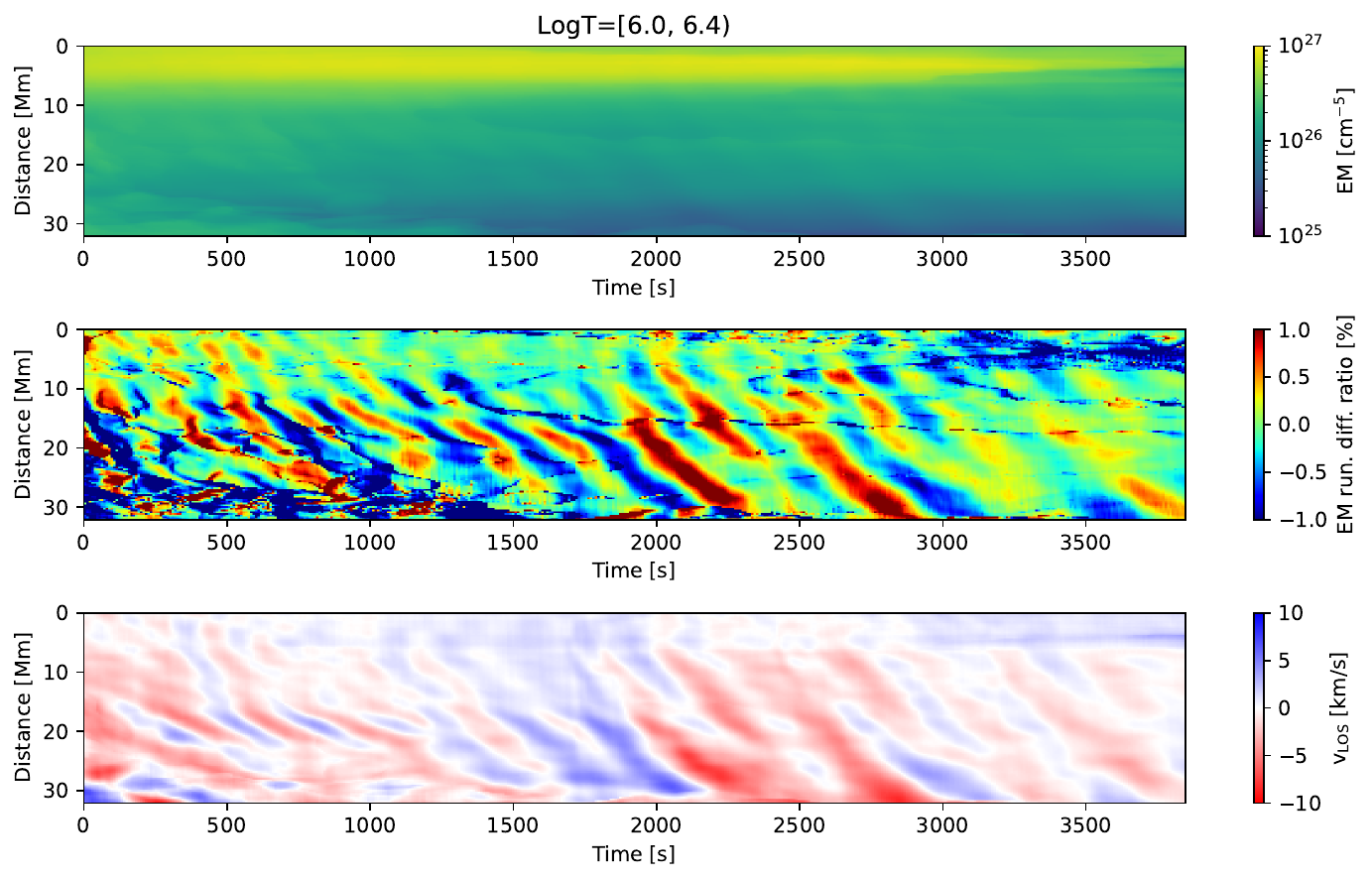}
\caption{Time-distance diagram illustrating the propagating disturbances along the slit shown in \figref{fig:doppler}. The upper panel displays the emission measure (EM) summed over the temperature between $10^{6.0}$ and $10^{6.4}$\,K. The middle panel shows the running difference ratio of the EM, which is the difference in the slit intensity at two consecutive times divided by that of the former. The bottom panel shows the Doppler velocity in the same temperature range, as shown in \figref{fig:doppler}.
\label{fig:wave}}
\end{figure*}

\begin{figure*}[t!]
\includegraphics[width=18cm]{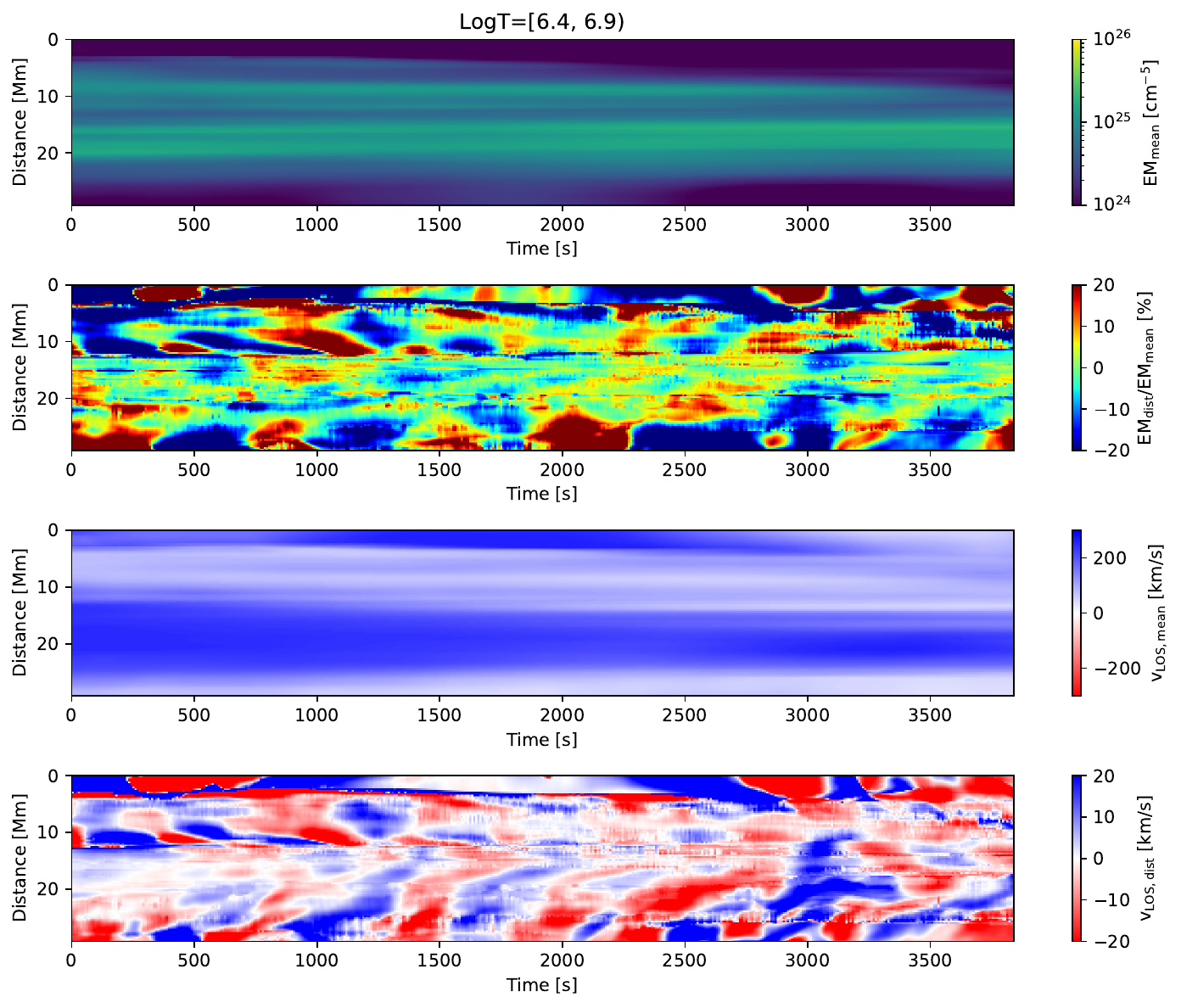}
\caption{Time-distance diagram along the slit shown in \figref{fig:doppler_up}. The top two panels display the mean  (EM$_{\mathbf{mean}}$) and disturbance  (EM$_{\mathbf{dist}}$) components of the emission measure, as described in \sectref{sec:doppler}. The bottom two panels show the decomposition of the Doppler velocity in the same fashion.
\label{fig:wave_up}}
\end{figure*}

In addition to imaging observations that provide a straightforward view of the dynamics of coronal EUV structures in the plane of the sky, Doppler shifts of spectral lines are the primary approach for detecting the line-of-sight motions of the coronal plasma. The data-driven MURaM simulations follow the standard output of the MURaM code and store the velocity along the 3 axes as a function of temperature in the range from $\log_{10}T=4.5$ to the highest temperature in the domain with an interval of $\Delta\log_{10}T=0.1$, as described in \citet{Chen+al:2022}. Therefore, these data facilitate visualizing the line-of-sight velocities of plasma at different temperatures or calculating spectral line profiles that can be compared with spectroscopic observations, such as data production from the upcoming Multi-slit Solar Explorer (MUSE) mission \citep{DePontieu+al:2022,Cheung+al:2022}.

\subsubsection{Doppler maps of coronal temperatures}
In this paper, we present a quick overview of the plasma dynamics in typical temperature ranges. The line of sight chosen here is along the $z$ axis. The emission-measure-weighted line-of-sight velocity $v_{\rm LOS}$ in the temperature range $[T_{\rm L}, T_{\rm H}]$ can be estimated by 
\begin{equation}\label{equ:emwv}
v_{\rm LOS} = \frac{\sum\limits_{T_{i}=T_{\rm L} }^{T_{\rm H}} v_{z}(\log T_{i}){\rm EM}_{z}(\log T_{i})}{\sum\limits_{T_{i}=T_{\rm L} }^{T_{\rm H}} {\rm EM}_{z}(\log T_{i})},
\end{equation}
where EM$_{z}(\log T_{i})$ and $v_{\rm z}(\log T_{i})$ are the emission measure and velocity in narrow temperature bins. $v_{\rm LOS}$ serves as a quick inspection at the Doppler velocity that can be observed in a spectral forming in the temperature range between $T_{\rm L}$ and $T_{\rm H}$. \figref{fig:doppler} shows the $v_{\rm LOS}$ of the $\Omega3$ model on Day 2. The full temperature range of the coronal plasma is divided into 4 bins to illustrate the characteristic dynamics of the plasma from the upper chromosphere to the transition region and from the warm to hot corona. We note that a higher temperature does not necessarily indicate a greater height in the domain and vice versa. The low and high temperature bins capture different components of the coronal plasma.

The lowest temperature bin in \figref{fig:doppler} demonstrates prevalent downflows in the entire domain, which is consistent with the fundamental facts found in observations\citep[e.g., ][]{Chae+al:1998,Peter+Judge:1999,Winebarger+al:2013}. The velocity field reveals a mixture of quiet Sun and low lying loops. 

In hotter bins, upflows start to appear, and their filling factor becomes more or less equal to that of downflows. The velocity field clearly depicts loop structures in the coronal volume. The strongest upflow is found immediately above the leading sunspot.

This trend continues at the typical coronal temperatures above $10^{6}$\,K. We can see prominent upflows that spread out from the leading sunspot, which maps the outbound loops that fan out. This eventually contributes to an upflow in the open funnel. An impact of the limited domain size and boundary is that part of the prevailing outflows from the leading spot propagates to the trailing spot through connections across the periodic boundary and becomes a siphon flow; otherwise, we could very likely see up-/outflows to both sides of the active region, which would be more consistent with previous spectroscopic observations \citep{Sakao+al:2007,Harra+al:2008,Warren+al:2011}. 

The hottest and most dynamic plasma in the model is found above the leading sunspot. We note that the hottest bin in \figref{fig:doppler} is displayed on a scale that is 5 times larger than those of the other panels. Therefore, we observe fast upflows of more than 100\,km/s along the open field lines connecting to the top boundary of the domain. 

To complement to the D2\_$\Omega$3 model shown in \figref{fig:doppler}, we present the emission-measure-weighted velocity of a model with an upflow top boundary (D2\_$\Omega$3\_Upflow run in \tabref{tab:list}) in \figref{fig:doppler_up}. The upper two panels show velocity maps that are highly consistent with those in \figref{fig:doppler}, suggesting that the plasma dynamics in closed loops are not affected by the top boundary.

In the panel for temperatures between $10^{6.0}$ and $10^{6.4}$\,K, while the velocities along closed loops remain similar to those in \figref{fig:doppler}, the upflows (outflows) above the leading sunspot are much higher. The most obvious contrast between the two cases is in the hottest bins. The upflow region in the lower right panel of \figref{fig:doppler_up} is greatly enlarged and has a much higher velocity than 200 km s$^{-1}$. This could contribute to the solar wind if these outflows can continue this tends; however, validating this possibility requires a much higher (and wider as well) domain \citep[e.g., as done by ][]{Iijima+al:2023}, which is far beyond the scope of this study.

\subsubsection{Waves in closed and open magnetic structures}
The overall pattern of the Doppler velocities in the active region remains stable during the 2 hours evolution time of a radiative MHD model (skipping the first hour). Interestingly, the time series of the Doppler map reveals many plasma dynamics on shorter time scales. We use an approach that is typically employed in observations to illustrate the periodic dynamics in the model corona. A slit along the $y$ direction is placed to the south of the active region core at $x=240$ and cuts through the loop group connecting the two sunspots. The time series of the emission measure, as well as its running difference ratio, and Doppler velocity along the slit for temperatures between $10^{6.0}$ and $10^{6.4}$\,K are displayed by the time-distance diagram shown in \figref{fig:wave}.

The time series of the emission measure, which represents the intensity of a spectral line forming in this temperature range, essentially shows how stable the coronal structures are during this hour. The weak periodic signal can be revealed only in the running difference ratio shown in the middle panel of \figref{fig:wave}. The difference in the slit intensity at two consecutive times is divided by the intensity of the former. This highlights relative intensity disturbances of a few percent. The slopes of the inclined ridges in the diagram correspond to speeds close to 50\,km/s along the slit in the plane of view. Given the slit position, these propagating disturbances, which are compressional, travel mainly across the loop top, i.e., transverse to the magnetic field. 

Similar ridge-like structures are also found in the time-distance diagram of the Doppler velocity, illustrating a transverse velocity disturbance with an amplitude of a few km/s. The intensity and velocity disturbances do slightly decay but remain clearly detectable during the 1 hour time displayed here. If such a loop is observed from a side view, the loop displacement in the $z$-direction is approximately 500 km, which appears to fall within the range of recent observations of decayless oscillations in coronal loops connecting sunspots \citep[e.g., ][]{Mandal+al:2022}.

For the model with an upflow top boundary (D2\_$\Omega$3\_Upflow), we place the slit across the hot and fast upflow region above the leading sunspot (lower right panel in \figref{fig:doppler_up}). The time-distance diagram is shown in \figref{fig:wave_up}. To enhance the wave disturbances in the intensity and velocity, we decompose the emission measure and Doppler velocity into a mean and disturbance component as follows.

The mean component of a quantity $\phi$ as a function of time (t) reads
\begin{equation}
\phi_{\mathrm{mean}}(t) = \frac{1}{\tau}\int_{-\tau/2}^{+\tau/2}\phi(t+t^{\prime})\mathrm{d}t^{\prime},
\end{equation}
where $\tau$ is the time-window for the moving average. The disturbance component is given by 
\begin{equation}
\phi_{\mathrm{dist}} = \phi - \phi_{\mathrm{mean}}.
\end{equation}
Here, we use $\tau=1000$\,s. Smaller values are also explored, and the illustration remains similar. The values of the mean and disturbance components obviously depend on window $\tau$; however, their sum remains exactly the original $\phi$.

The top two panels in \figref{fig:wave_up} reveal compressive propagating disturbances, with an EM disturbance of more than 10\% of the mean values. The lower panels illustrate a steady upflow of 200 km s$^{-1}$ and a velocity disturbance of 10 - 20 km s$^{-1}$. Given the highly vertical magnetic field in this region (10$^{\circ}$ or less from the $z$ axis), the line-of-sight velocity largely shows longitudinal disturbances.

A detailed analysis of the waves in the simulations presented here is not the focus of this paper and is left for future investigations. What triggers the transverse and slow propagating waves in a low-$\beta$ corona remains an interesting question. Here we choose to present results in a way similar to how actual observations are made, where wave dynamics in a single loop may be strongly contaminated by line-of-sight integration. More in-depth analysis needs to isolate the properties of the wave disturbances and oscillations along a particular loop in the 3D volume, which then can then be compared with those of classical loop models. We expect to investigate whether this is similar to the model of \citet{GaoYuhang+al:2023} where small amplitude oscillations are driven by p-modes. Meanwhile, although projections of transverse oscillation might not be completely ruled out, the compressive longitudinal disturbances in the open flux region are more consistent with slow-mode magnetoacoustic waves.

\subsection{The Corona Heating beneath the Observables}

\begin{figure*}[t!]
\includegraphics[width=18cm]{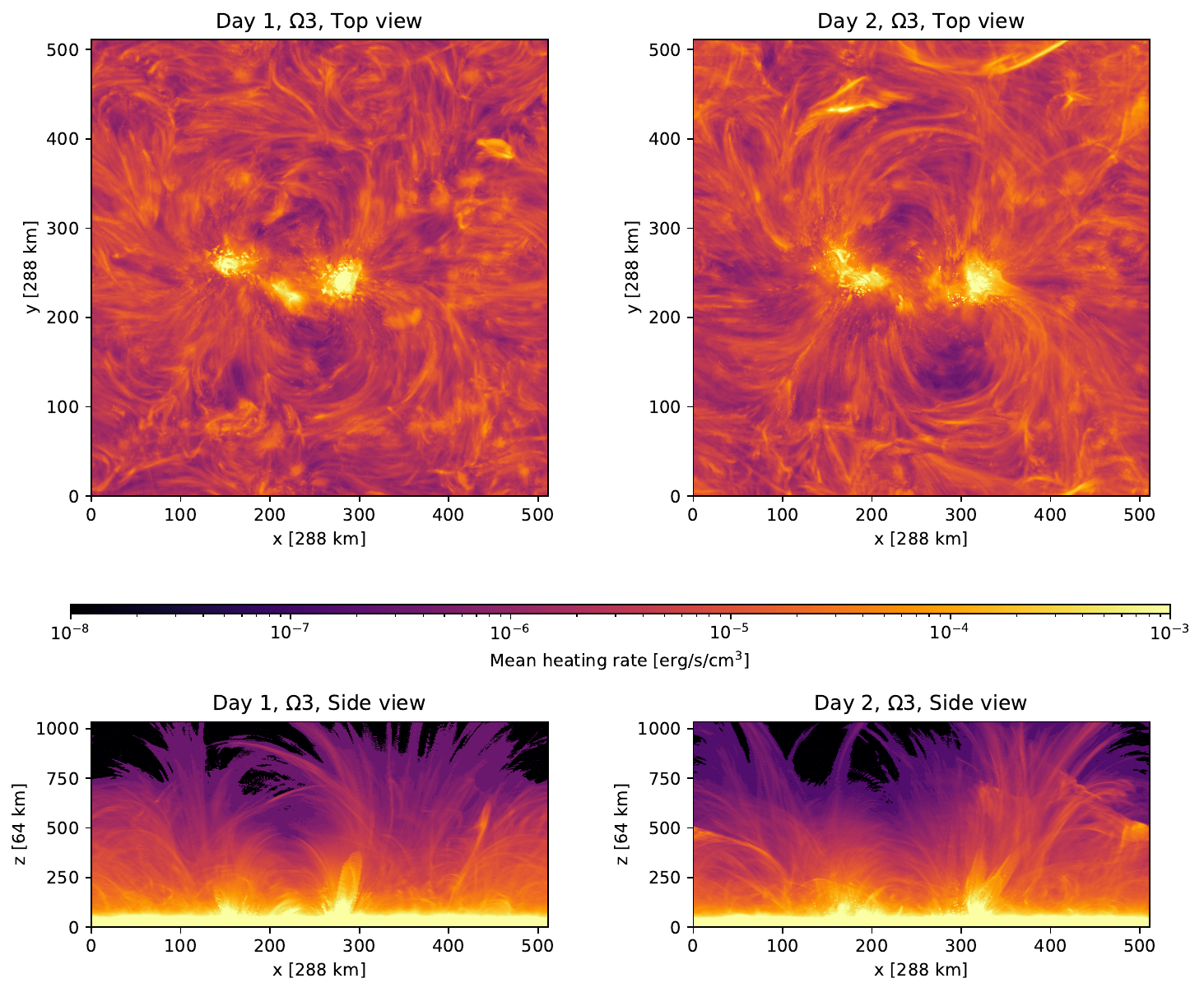}
\caption{The mean coronal heating rates of the $\Omega3$ models on Day 1 (left column) and Day 2 (right column). The top view presents an average along the $z$ direction between the coronal base and the top of the domain. The side view presents an average of the $y$ direction between $y=50$ and 450. The points close to the $y$ boundary are excluded to avoid an unrealistic magnetic separatrix due to the periodicity.    
\label{fig:q3d}}
\end{figure*}

\begin{figure*}[t!]
\includegraphics[width=18cm]{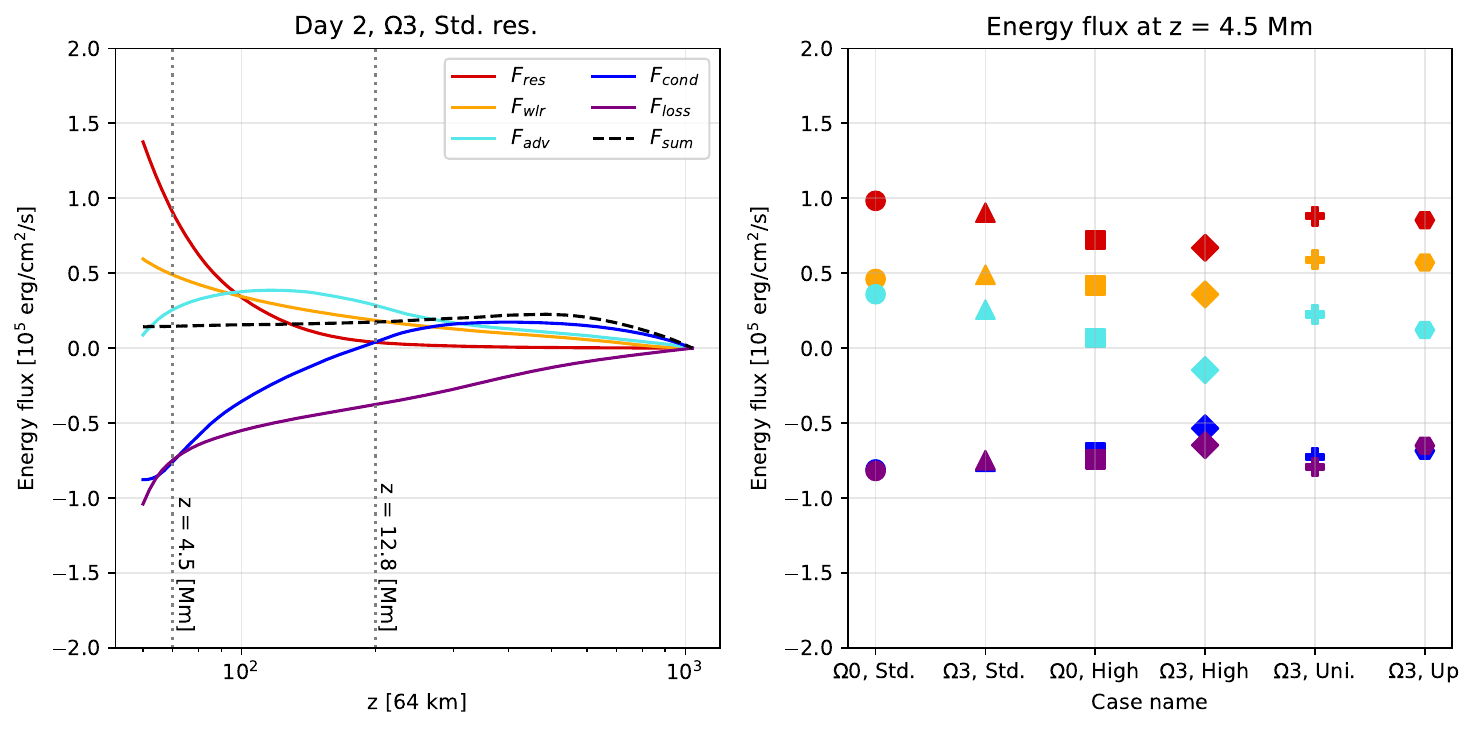}
\caption{The vertical energy fluxes that are defined by \equref{equ:flux} in the main text and depict the energy transport and dissipation through the vertical domain. In the left panel, the energy fluxes for resistive heating ($F_{\rm res}$), Lorentz force work ($F_{\rm wlr}$), thermal conduction ($F_{\rm cond}$), optically thin radiative loss ($F_{\rm loss}$), and advection ($F_{\rm adv}$), which includes enthalpy and kinetic energy fluxes are plotted, as indicated by the legend. The black dashed line represents the sum of the 5 terms. The axis for the height $z$ shows in the number of vertical grid points on a logarithmic scale to highlight the lower part of the domain where the quantities change rapidly. The right panel compares the energy fluxes at the coronal base (as indicated by the vertical dotted line) for the 6 models on Day 2. The quantities are shown in the same colors as those in the left panel.
\label{fig:heating}}
\end{figure*}

\begin{figure*}[t!]
\includegraphics[width=18cm]{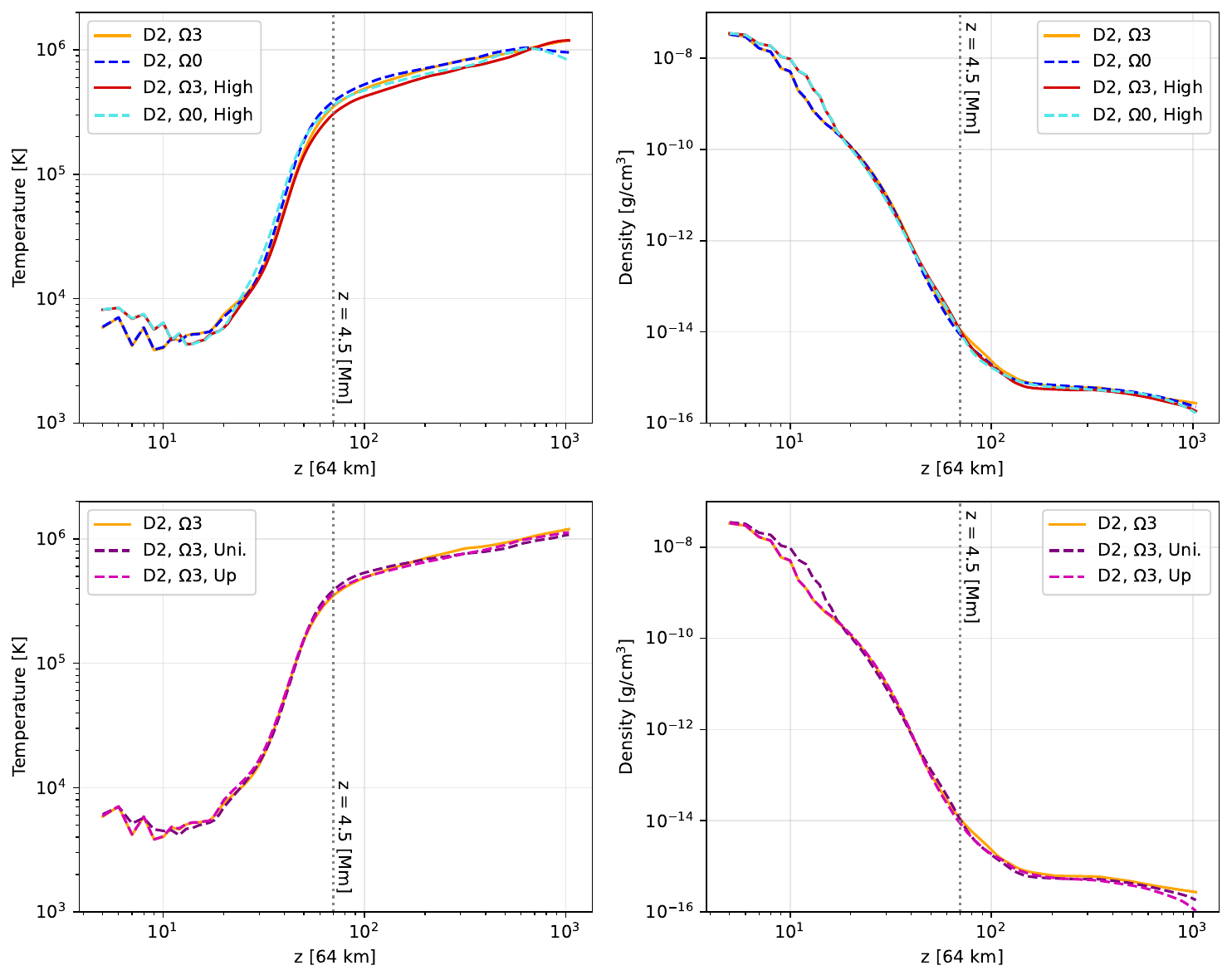}
\caption{Coronal density and temperature as a function of height. The results from the 6 models compared in the right panel of \figref{fig:heating} are plotted, as indicated by the legend. The horizontal and temporal averaging is performed in the same manner as in \figref{fig:rhot_z}. The Day2\_$\Omega$3 model is plotted in both rows as a reference.
\label{fig:rhot_sup}}
\end{figure*}

\begin{figure}
\includegraphics[width=8.5cm]{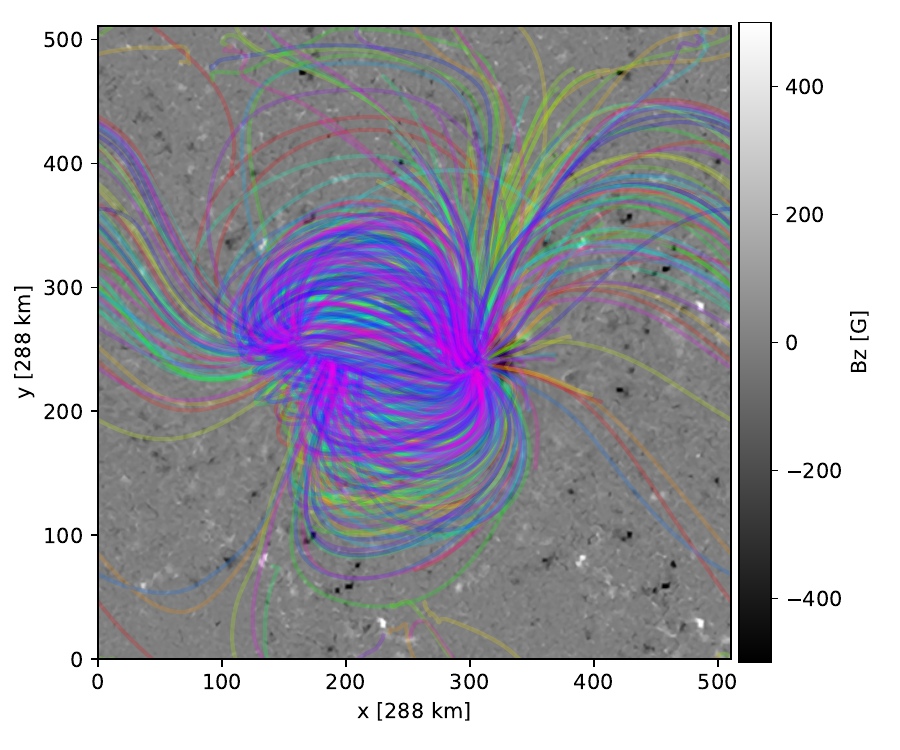}
\caption{Magnetic field lines for the analysis of coronal heating shown in \figsref{fig:qzs} and \ref{fig:qbl}. The grayscale image shows $B_{z}$ in the photosphere ($z=0$). Field lines are calculated from 1000 seed points that are randomly distributed in the center part of the coronal volume (see main text for details). Line colors indicate line indices. A line passing through the periodic side boundary, which appears as two segments, is treated as a complete line in the analysis.
\label{fig:lines}}
\end{figure}

\begin{figure*}[t!]
\center
\includegraphics[width=18cm]{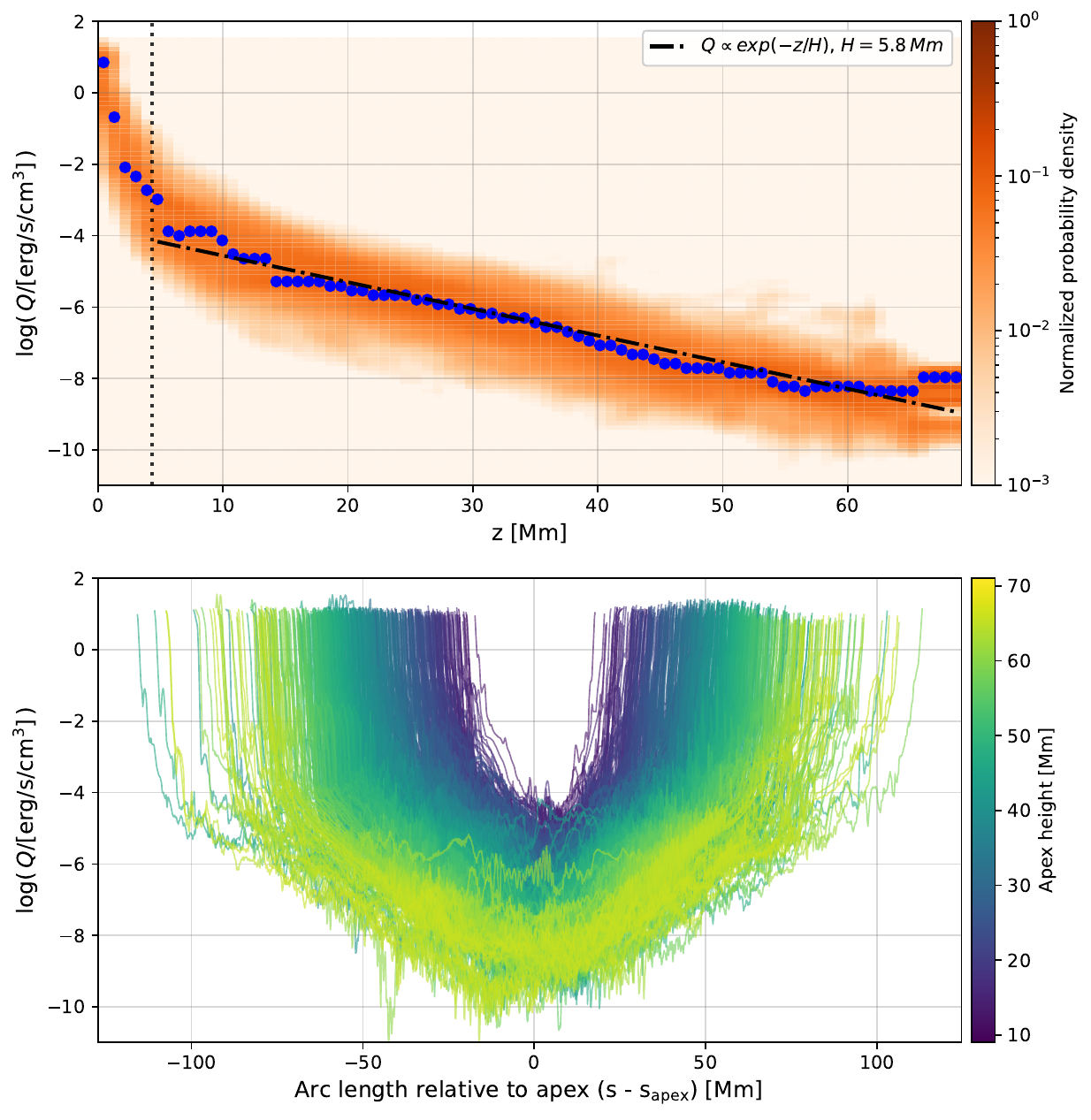}
\caption{The distribution of the volumetric coronal heating rate ($Q$, on a logarithmic scale) with height ($z$) and arc length ($s$) along the field lines. Only closed field lines are considered. Top panel: Normalized joint probability distribution function of $Q$ and $z$. The blue dots indicate the most probable values of $Q$ at each $z$. The black dash-dotted line shows a fitting of the most probable value of $Q$ in the corona ($z\geq4.5$ Mm, as indicated by the vertical dotted line) by an exponential function. Bottom panel: $Q$ as a function $s$. Different field lines are aligned by their apex and displayed in colors corresponding to the height of their apex points.    
\label{fig:qzs}}
\end{figure*}

\begin{figure*}[t!]
\includegraphics[width=18cm]{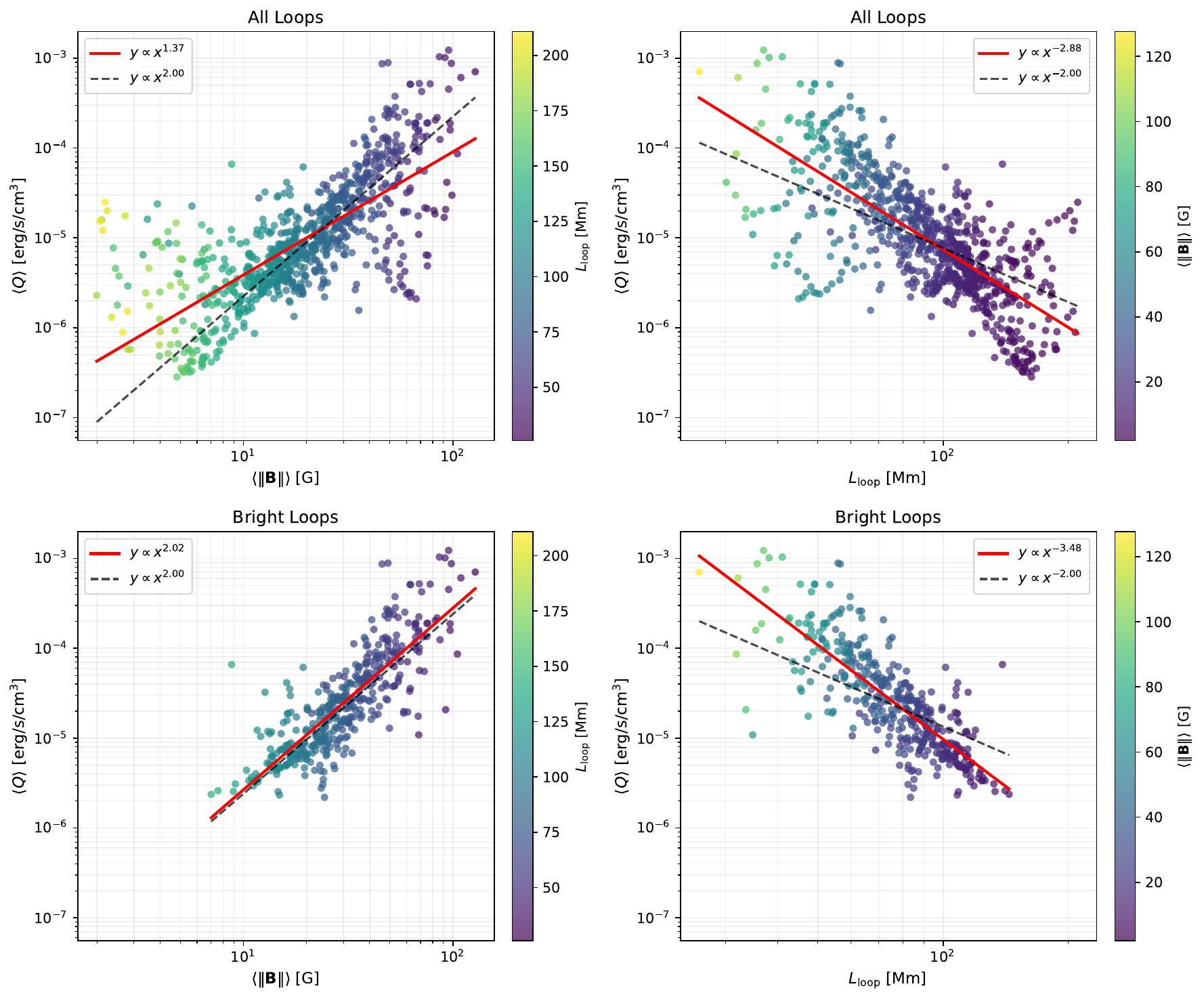}
\caption{Relation between the mean volumetric heating rate ($\langle Q \rangle$) and the mean magnetic field strength ($\langle \Vert\mathbf{B}\Vert \rangle$) along field lines and the relation between $\langle Q \rangle$ and the loop length ($L$, i.e., the length of field lines). Top row: All closed field lines (loops). bottom row: Bright loops that are the top 50\% in the distribution of $\log(\langle n_{e}^{2} \rangle)$ (see main texts for details). Left panel: The red solid line shows the fitting of the data points with a power function. The black dashed line indicates a power index of 2.0 for reference. The data points are color coded according to the loop length. Right panel: The red solid line shows the fitting of the data points with a power function. The black dashed line indicates a power index of -2.0 for reference. The data points are color coded by the mean magnetic field strength.
\label{fig:qbl}}
\end{figure*}

All the observed coronal structures and their dynamics are manifestations of energy deposition in the corona and thus can be used to constrain the properties of coronal heating. It may be argued that the smallest scales revolvable by current MHD models remain much larger than those where heating of the real corona occurs (e.g., the size of a nanoflare or the thickness of current sheets predicted via theoretical dissipation coefficients). Braiding models performed with a resolution that is much better than that of current instruments or active-region scale models \citep{Breu+al:2022,Breu+al:2024} tested the same coronal heating mechanism as in the large-scale models and obtained consistent coronal plasma properties. Therefore, realistic 3D MHD coronal models that can reproduce fundamental observational properties of actual active regions on the Sun, such as those presented in this paper and in previous works, are able to provide valuable insight into the heating in the coronal volume, at least down to the scales that they can resolve.

\subsubsection{Heating rate in 3D space}\label{sec:q3d}
The coronal heating rate in the MURaM code is contributed by the sum of the viscous dissipation of the kinetic energy and the resistive dissipation of the magnetic energy. The former is automatically added to the internal energy because the code solves the conservation of the sum of internal and kinetic energy, whereas the latter is explicitly added into the energy equation. We present the mean volumetric coronal heating rates in the $\Omega3$ models on Day 1 and Day 2 in \figref{fig:q3d}. A line-of-sight average is performed from a top view and a side view, respectively. Naturally, the heating rate in the lower atmosphere is orders of magnitude greater than that in the corona. Therefore, when calculating the average from the top view, we adopt only the domain above the coronal base ($\sim$5\,Mm). For the average from the side view, the first and last 50 grid points next to the boundary in the $y$ direction are excluded, because a magnetic separatrix builds up at the periodic boundary and gives rise to some current structures that should not exist in the real active region.

The strongest heating rate on Day 1 are found above the two sunspots and in a patch between them. Although we have applied a factor $f_{v}$ in the initial velocity field to avoid artificial disturbances in very strong magnetic fields, braiding is still at work for magnetic fields of kilo Gauss, which can effectively generate a high energy flux in sunspot areas. Moreover, the observation that drives the simulation already includes the evolution of the sunspots, which also leads to a significant energy flux bound to the sunspot areas, as we found when driving a coronal model with a magnetoconvection simulation of active region emergence \citep{Chen+al:2015}. Compared with the magnetogram shown in \figref{fig:mag}, the high heating between the two sunspots on Day 1 is cospatial with a positive-polarity magnetic patch intruding into an area close to the negative-polarity flux concentrations. It is easier to build stronger current sheets (on scales resolvable by our resolution) that can yield both high viscous and resistive heating. Observations have revealed heating events and magnetic reconnection in the very early stage of flux emergence \citep[e.g., ][]{Toriumi+al:2017,TianHui+al:2018}, which is a similar scenario at smaller spatial scales.

The side view of the mean heating rate on Day 1 shows that the patch with the strong heating between the sunspots appears in the shape of lying loops. The enhanced heating above the sunspots, particularly above the leading sunspot, is more extensive in height. In the higher corona, the most evident structures of the heating rate appear to be numerous thin threads that may form loop-like arches or open funnels. These structures are likely manifestations of coronal current sheets that build up in response to forced driving at the bottom boundary, and they demonstrate similar shapes as the observed EUV loops in this active region. Although this appearance of the current structures motivated indirect comparisons that have been extensively used in previous studies, the similarity between the model and the reality can be assessed in a more exact manner only by comparing the same observable quantities, as shown in the earlier sections.

On Day 2, more loop-like structures connecting the sunspots become visible in the top view of the heating rate. The heating in the patch between the two sunspots significantly decreased, probably because the magnetic concentrations of different polarities are more separated after one day of emergence. The side view of the heating rate on Day 2 looks similar to that on Day 1, but the heating rate, especially for the thread-like structures in the coronal volume, becomes noticeably stronger. This heating distribution on Day 2 generates the plasma density and temperature structures displayed in \figref{fig:rhot} (the $\Omega3$ model) and eventually the EUV image shown in \figref{fig:dive} (the $\Omega3$ model).

\subsubsection{Coronal energy balance}
Establishing a stable and hot corona essentially means a balance between the heating and the energy loss through optically thin radiation and thermal conduction. In an evolving corona, this may include an advective flux and works done by external forces. Here, we investigate the coronal energy balance, more precisely speaking, the balance of the energy equation solved in MURaM simulations, in different model setups.

Because the domain is periodic in the horizontal direction, only the vertical flux matters in the transport of energy through the domain. As in our previous study \citep{Chen+al:2022}, the conservative form of the energy equation is integrated over volume and divided by the area of the horizontal domain. The advection and conduction terms are genuine fluxes. A volumetric energy change rate $Q$, for example, a heating rate or radiative loss rate, is reformed to energy flux by
\begin{equation}\label{equ:flux}
F(z) = \frac{1}{L_{x}L_{y}}\int_{z}^{z_{\rm top}}Q\,dxdydz,
\end{equation}
such that the integral of $Q$ in the volume from a certain height $z$ to the top of the domain\footnote{In practice, the top 2\% of grid points, where the vertical energy fluxes are nevertheless vanishing, are omitted.} is identically represented as the vertical flux $F(z)$ at the bottom boundary of the domain. This is done for the resistive heating $F_{\rm res}$, Lorentz force work $F_{\rm wlr}$, and optically thin radiative loss $F_{\rm loss}$. The advective flux $F_{\rm adv}$ and thermal conduction flux $F_{\rm cond}$ are averaged in the horizontal direction. These fluxes are also averaged over the same time period for the mean density and temperature profiles shown in \figref{fig:rhot_z}. The 5 most important vertical energy fluxes that determine the evolution of the plasma energy (i.e., the sum of the internal and kinetic energy) are plotted in \figref{fig:heating}. The work done by gravity is omitted, as it is much smaller than the other terms. An alternative is to show the volumetric energy change rate and the divergence of an energy flux \citep[as in ][]{Rempel:2017}, which essentially presents the same information as the flux formulation shown here. Similar to \figref{fig:rhot_z}, the height axis is plotted as the number of vertical grid points on a logarithmic scale such that the lower atmosphere ($z>3.84$\,Mm is shown in this plot) is greatly stretched while the corona part is compressed. The two vertical dashed lines indicate marks of the actual height; in particular, $z=4.5$\,Mm is about the coronal base.

At the coronal base, the primary terms in the balance are the resistive heating versus the energy loss through optically thin radiation and thermal conduction, with the rest compensated by the Lorentz force work and upward energy flux by advection. The Lorentz force work directly contributes to the kinetic energy, which can be partly dissipated to heat the plasma. Previous MURaM coronal simulations \citep{Rempel:2017,Chen+al:2022} have consistently demonstrated that the viscous heating, which is closely correlated with the Lorentz force work in a corona in dynamic equilibrium, contributes more coronal heating than the resistive heating does, in a high magnetic Prandtl number setup that is likely more consistent with the condition of the real corona. The simulations shown in this paper use a high Prandtl number setup identical to that used in the previous MURaM simulations; however, the resistive heating is predominant in the lower part of the domain ($z<6.4$\,Mm, as shown in \figref{fig:heating}). This is caused by the driven evolution of the magnetic field that induces stronger current, particularly in regions with strong magnetic field. Although \figref{fig:q3d} shows the sum of viscous and resistive heating, the strong heating regions discussed in \sectref{sec:q3d} (e.g., those cospatial with the sunspots) are from the resistive heating. The resistive heating decays much faster with height, and the viscous heating prevails in the middle and higher parts of the domain.

As expected, the cooling by thermal conduction peaks at the coronal base (or the top of the transition region), where a high temperature is coupled with a steep gradient. Moreover, the thermal conduction in a large part of the corona ($z>13$\,Mm) actually contributes to a heating instead of cooling, which means that, on average, the temperature peak is in the lower corona. Similar behaviors was also found in other MURaM simulations, for example, in the quiet Sun case shown in \citet{Chen+al:2022}. This is a result of a fundamental property of coronal heating and density stratification (at least in the 3D coronal models): the average heating rate decays with height, whereas the plasma density decreases even faster. Therefore, a peak of the heating per mass, which is proportional to the temperature increase, is found in the lower corona, as demonstrated by \citet{Bingert+Peter:2011}. Similarly, the model of \citet{Hansteen+al:2015} also indicates a mean temperature peak in the lower corona.

The major terms in the energy equation at the coronal base are compared for the 6 models on Day 2, as plotted in the right panel of \figref{fig:heating}. The energy fluxes are shown in the same color as in the left panel and are lined up by their case names, as marked on the horizontal axis and by corresponding symbols. Resistive heating is the major heating source in all the models. The total amount of heating in all the models with the standard resolution is virtually the same and does not depend on the choice of $\Omega$ or the initial condition of the plasma. In comparison, the models with the high resolution have approximately 30\% smaller heating inputs, which is consistent with the fact that they yield lower synthetic EUV intensities. This decrease in heating is not too severe because we have seen in the results shown in earlier sections that the high resolution models still present coronal structures that are similar to the observation of this active region. Moreover, the horizontally and temporally averaged plasma density and temperature for the 6 models compared here are plotted in \figref{fig:rhot_sup} in the same fashion as in \figref{fig:rhot_z}. The mean density and temperature profiles demonstrate that despite small differences in the coronal temperatures, all the models are able to reproduce a stratified solar atmosphere that comprises a cool and dense chromosphere, a transition region with a steep temperature gradient, and a hot and tenuous corona. It is also interesting to note that the model with an upflow top boundary yields a clearly lower density in the higher part of the domain, which is an expected result of the persistent outflows along open field lines.

\subsubsection{Heating in coronal magnetic loops}

We randomly select 1000 magnetic field lines in the central part of the domain to analyze the distribution of the coronal heating rate in the magnetic field. The horizontal positions of the seed points are distributed within the range of $[\frac{1}{3}L_{x},\frac{2}{3}L_{x}]$ and $[\frac{1}{3}L_{y},\frac{2}{3}L_{y}]$, which covers the main sunspots in the active region, and the heights of the seed points are distributed in the range of $[\frac{1}{6}L_{z}, \frac{2}{3}L_{z}]$ (approximately [10\,Mm, 50\,Mm]). These field lines are displayed in \figref{fig:lines} above a magnetogram. In the following, we use only closed field lines, which have both footpoints in the photosphere (instead of the top boundary). Thus, field lines that pass through the periodic side boundaries are identified as a complete line and a valid closed field line.

We calculate the joint probability distribution function between the volumetric heating rate ($Q$), which is the sum of the viscous heating ($Q_{\rm vis}$) and resistive heating ($Q_{\rm res}$), and the height ($z$) for all the points along all the field lines. The bin size for $z$ is 0.864\,Mm and the bin size for $\log Q$ is 0.2. The upper panel of \figref{fig:qzs} presents the normalized probability density. For each bin of $z$, we retrieve the most probable values of $Q$, which are marked in \figref{fig:qzs} as blue dots. The heating rate decays rapidly with height, particularly in the first few Mm. Above the coronal bottom ($z=4.5$\,Mm), the most probable values of $Q$ decreases exponentially with scale height (H) of 5.8\,Mm. Similar behaviors have been previously reported in other models \citep[Figure 7 in ][]{Gudiksen+Nordlund:2005a}. \citet{Bingert+Peter:2011} found that the horizontally averaged heating rate above 4\,Mm in their model decays exponentially with a scale height of 5\,Mm. Both values are similar to what we find here. It is intriguing to note that two models consider totally independent active regions and differ significantly in domain size. 

The lower panel of \figref{fig:qzs} displays the distribution of $Q$ along the loop arc ($s$), i.e., magnetic field lines, similar to what has been shown in Figure 6 in \citet{ShiTong+al:2024}. All the lines are aligned by their apex and color-coded according to their apex height. The photospheric footpoints of all the lines have rather high heating because of the strong stresses in the high plasma $\beta$ regime. Similar to the trend revealed in the upper panel of \figref{fig:qzs}, $Q$ quickly decreases along the loop arc (moving toward $s-s_{\rm apex}=0$) and decays more or less exponentially with $s$ for the majority of the loop.

The footpoints (and lower segment) of the loops in our model have a much higher heating rate than those shown in \citet{ShiTong+al:2024}, although their model also places the bottom boundary at the photosphere. This is likely because we have a better horizontal resolution (288 km vs. 4.5\,Mm) to resolve small scale velocities at the solar surface. Moreover, the heating rate in our model decays much faster than that in the model of \citet{ShiTong+al:2024}. For example, the heating scale height in their model is generally larger than 150\,Mm. This reflects the difference embedded in the nature of the assumed heating mechanism. The braiding of the magnetic field builds stresses and heating that are more concentrated at the footpoints, whereas the alfv\'enic waves/turbulences transport energy to much greater heights.

We analyze the scaling relations between the heating rate and fundamental parameters of the coronal loops. A detailed investigation is vital but beyond the scope of this study. Here we focus on the magnetic field strength $\Vert\mathbf{B}\Vert$ and full loop length $L_{\rm loop}$.

Given the conservation of flux along a magnetic tube, the cross section ($A_{\rm loop}$) of a coronal magnetic loop is proportional to $1/\Vert\mathbf{B}\Vert$. The mean volumetric heating rate ($\langle Q\rangle$) in a coronal loop is thus estimated by
\begin{equation}\label{equ:qmean}
\langle Q \rangle = \frac{\int A_{\rm loop} Q \mathrm{d}s}{\int A_{\rm loop}\mathrm{d}s}.
\end{equation}
The mean magnetic field strength ($\langle\Vert\mathbf{B}\Vert\rangle$) is given by  
\begin{equation}\label{equ:bmean}
\langle\Vert\mathbf{B}\Vert\rangle = \frac{\int A_{\rm loop} \Vert\mathbf{B}\Vert \mathrm{d}s}{\int A_{\rm loop}\mathrm{d}s} = \frac{\int \mathrm{d}s}{\int 1/\Vert\mathbf{B}\Vert\mathrm{d}s}.
\end{equation}
All calculations consider the closed field lines and only the coronal segment above the coronal bottom ($z \geq 4.5$\,Mm).

The relation between the mean volumetric heating rate and loop parameters are plotted in \figref{fig:qbl}. The upper left panel indicates that the majority of the data points follow a power relation between $\langle Q\rangle$ and $\langle\Vert\mathbf{B}\Vert\rangle$, whereas a few data points scatter at two wings of rather small or large mean magnetic strength. The fitting to all the data points yields a power index of 1.37, whereas the trunk of the distribution seems to follow a power index of 2, as indicated by the dashed line in the upper left panel of \figref{fig:qbl}. Such an index is proposed by theoretical scaling relations based on scenarios of stochastic braiding or magnetic current sheets, as discussed by \citet{Mandrini+al:2000}. 

We perform a further filter to select the bright loops in the data points. The average electron number density squared ($\langle n_{e}^2 \rangle$) is estimated by
\begin{equation}
\langle n_{e}^{2} \rangle = \frac{\int A_{\rm loop}\,n_{e}^{2}(\log T\ge5.3)\,\mathrm{d}s}{\int A_{\rm loop}\mathrm{d}s}.
\end{equation}
This quantity serves as a proxy of the temperature integrated emission measure (albeit without the line of sight effect). The top 50\% of the loops, which corresponds to a mean number density larger than 10$^{8.55}$cm$^{-3}$, are shown in the lower left of \figref{fig:qbl}, and the power index of 2.02 well matches the theoretical relation.

The upper right panel of \figref{fig:qbl} reveals a similar distribution in which the trunk clearly indicates a power relation, whereas the scatter points reflect short loops with insufficient heating and a few strongly heated long loops. The dashed line corresponds to a power index of -2, which is derived from observed soft X-ray loops \citep{Porter+Klimchuk:1995}, and implies that the heating is concentrated at shorter loops. The fitting to the data points yields a power index of -2.88. The lower right panel considers the data points from the bright loops, which constitute the trunk of the distribution, and a power index of -3.48 fits the trend of the trunk. This means that the heating in the model is severely footpoint-concentrated and decays very quickly in height (length), which is in line with the fact that the long loops and diffusive corona in the higher part of the domain are insufficiently heated, as already revealed in the comparison with observations.

\section{Discussion and Conclusion}\label{sec:conclusion}
In this paper, we present the application of the data-driven MURaM code to AR11640, which follows the emergence of this active region across 4 solar days. The radiative MHD model on each day reproduces the development of brighter and longer coronal loops as the active region emerges, although the detailed geometry of some loops is affected by the connectivity through the periodic horizontal boundary utilized in the simulation. The model reveals the fine structures of the coronal heating in the coronal volume and how the heating evolves and gives rise to a hotter and loop-dominated corona, as the active region emerges. Although no spectroscopic observations are available for this active region, we present line-of-sight velocities in wide temperature ranges that reveal abundant plasma dynamics and propagating waves in the model corona, as expected on the real Sun.

\subsection{The Free Parameter in the Bottom Boundary Driver}
The same method can generally be applied in other active regions, as long as a time series of the magnetic field (at least the vertical component) is available.This means that either the target active region is not very close to the limb in the time period of interest, or that a vector magnetic field measurement is available from multiple aspects away from the Sun-Earth direction. Although not conducted in this study due to our limited computational resources, it would be interesting to test the change in the results if the radial component derived from the line-of-sight magnetic field using the radial assumption is used. 

A primary free parameter in this model is the constant $\Omega$, which is multiplied by $B_{z}$ and provides a necessary constraint to calculate the horizontal electric field. The setup with a larger $\Omega$ adds a more strongly twisted horizontal magnetic field, whereas the vertical component remains unaffected and follows the observed magnetogram. This treatment is equivalent to adding a so-called noninductive component to the horizontal electric field to reconstruct the time series of the observed magnetic field. \citet{Cheung+DeRosa:2012} and \citet{Lumme+al:2022} demonstrated the role of this noninductive component in injecting free magnetic energy and forming magnetic flux ropes. In our study, only small $\Omega$ values are employed. As a result, the mean coronal plasma properties are similar in all the models with different values of $\Omega$. The effect of the $\Omega$ on the large scale corona is secondary to the change in the active region magnetic field due to flux emergence. The role of $\Omega$ is evident in particular coronal loops. We notice that a moderate value in this study ($\Omega3$ models) provides the best similarity between the model and observations. 

Nevertheless, the investigation of this parameter aims to understand how the models would change with different nonpotential magnetic fields imposed. We do not intend to infer a single parameter that can best fit the observation, nor do we expect that a single parameter that is constant in time and space can describe the various situations in real active regions. In the future, it would be interesting to compare the current results with those of models driven by the electric field obtained via, for example, the characteristic data-driven boundary developed by \citet{Tarr+al:2024} or the PDFI method \citep{Kazachenko+al:2014,Fisher+al:2020}, which has been applied to model active regions of major eruptions \citep[e.g., ][]{Afanasyev+al:2023}.

\subsection{The Impacts of the Horizontal and Top Boundaries}
A considerable difference between the numerical models and real active regions is the horizontal boundary. With a Cartesian domain covering a limited part of the solar disk, perhaps no boundary conditions can perfectly capture the complex connections between the target active region and the ambient quiet Sun and other active regions. As shown in \sectref{sec:result}, the periodic horizontal boundary allows field line connections through the boundary and hence prevents the formation of some large coronal loops that are expected to connect within the domain. 

The limited height extension of the domain also has a nonnegligible effect on the formation of coronal structures. For example, part of the open funnel could be loops that are closed at much larger spatial scales, which can obtain heating input from both footpoints. However, the heating input of the open funnel in the models is primarily obtained from the footpoint at the bottom boundary, and the side connected to the top boundary of the domain usually becomes a pure outflow. This limits the mass filling in the these funnels and leads to a lower EUV intensity. 

An obvious approach to mitigate the undesirable effects of the horizontal and top boundaries is to add extra padding in the horizontal direction and levitate the top of the domain, with the price of a substantial increase in computational expenses. Although not performed in this study, we expect to conduct this experiment when resources permit. At this stage, a static/adaptive mesh refinement technique that can greatly benefit a large domain and fine grid spacing is not available (not even a stretched mesh) in the MURaM code; we will also seek opportunities to implement physical processes such as done in the MURaM code in other frameworks with flexible meshes.

\subsection{How Should a Model Be Compared with Observations?}
A comparison between numerical models and observations is important in this study, as well as in all studies that aim to reproduce a particular active region or eruption via numerical simulations. The available output quantities depend on the assumptions used in the models. For example, only magnetic field information is meaningful in magneto-frictional models or zero-$\beta$ MHD models, whereas plasma thermodynamic properties are nonetheless available in MHD models even if they only solve an adiabatic or isothermal energy equation. For comparison with observations, emission proxies based on the current density squared, which assumes a relationship between the coronal emission and Ohmic heating, are often used in magnetic field models. Despite a simplified energy equation, all MHD models may easily generate synthetic EUV images by the plasma density and a temperature response function of a certain instrument and compare synthesized images with observations. Although this very often leads to the conclusion that a model is consistent with observations, such comparisons remain qualitative rather than quantitative. 

More realistic energy transport terms need to be considered, such that plasma thermodynamics and their evolution could be more consistent with those in the real corona. The model synthesized observables based on these plasma density and temperature properties are more meaningful for making a quantitative comparison with actual observations, as was done, for example, by \citet{Warnecke+Peter:2019}. They also demonstrated the difficulty of truly reproducing an observed active region quantitatively. Even many of the loops in the model appear to have similar shapes as those seen in the observed EUV images, \citet{Warnecke+Peter:2019} noted that the actual count rate given by the model is lower than the observation by a factor of 6, corresponding to a factor of approximately 2.5 in density. The intensity in our model is also lower than the actual observation but by no more than a factor of 2, which means a factor of approximately 1.4 in density. 

We have shown in \sectref{sec:result} that the coronal heating and coronal plasma properties present many more fine structures, which, via direct visible inspection, are very similar to the observed EUV structures of the real active region, particularly when a certain mask is applied and the dynamic range is fine-tuned. We also demonstrated that even in this case, the model synthesized emission, which is arguably the only quantity that can be directly compared with observations, is not necessarily consistent with the observed EUV images. It may be more often to see the opposite. The spatially and temporally varying heating rate dynamically changes the mass filling and temperatures of coronal loops in a way that may differ significantly from static situations such as the scaling law \citep{RTV}. Moreover, the long-known line-of-sight integration must also play a role in determining the final observable appearance of the active region corona \citep{Malanushenko+al:2022}. Thus, we suggest that all perfectly consistent qualitative comparisons need to be taken with a grid of salt.

We also note that similar to the model of \citet{Warnecke+Peter:2019}, our models do not generate enough hot plasma emission in the active region core as revealed by AIA 94 images. A similar result of missing the hottest plasma in active regions was also found in the global-scale model of \citet{ShiTong+al:2024}. We expect that, instead of a high resolution numerical simulation, a higher-resolution observation capturing the complex magnetic structures in the active region core may help to solve this issue. The model of \citet{Lu+al:2024} demonstrated how hot plasma in the active region core is sustained by continuous magnetic reconnections in a multipolar magnetic field configuration. This scenario can be scaled down for smaller magnetic flux concentrations and shorter loops and will be tested in our future work.

Another important aspect of the comparison is the plasma dynamics. This requires spectroscopic observation, which is unfortunately not available for the active region in this study. \citet{Bourdin+al:2013} reported consistent Doppler shift patterns in loops formed in the model and at the same locations in the observed Doppler map. However, that observation also revealed many more structures and dynamics in the intensity and Doppler maps than the model shows. A recent state-of-the-art observation by \citet{ZhuYingjie+al:2025} demonstrated various flow patterns in different parts of a decaying active region. It is intriguing to assess in future studies whether data-driven radiative MHD models can self-consistently reproduce these flows with a high resolution observation of the magnetic field of the target region.

\subsection{Can Active Region Models be Useful for Wave Studies?}
In general, waves are of great interest in studies of solar corona because of their potential role in transporting energy that may heat the corona and in diagnostics of the coronal plasma and magnetic field \citep{Nakariakov+al:2016,VanDoorsselaere+al:2020}. A solid theoretical basis has been established for classical straight magnetic flux tubes. However, only a few models exist that allow the properties of waves in a curved magnetic field resembling coronal loops to be studied \citep[see e.g, ][and references therein]{Ofman+Wang:2022,Lopin+Nagorny:2023,GuoMingzhe+al:2024,ShiMijie+al:2025}, or consider the coronal radiative loss and thermal conduction that are crucial to the evolution of the plasma thermodynamics and thus the waves \citep[e.g., ][]{Kolotkov+al:2020,VanDamme+al:2020,ShiMijie+al:2021,GuoMingzhe+al:2023}. Although compared with dedicated loop models, extracting wave signals and isolating the properties of a particular structure in a dynamically evolving active region model may become difficult, a clear benefit of the current and similar models \citep[e.g., ][]{Chen+Peter:2015} is a realistic magnetic configuration of an active region corona combined with self-consistently evolving plasma, which creates an environment closer to those where waves and oscillations are observed. 

New observations with more advanced instruments continue to reveal waves that have long been expected \citep{Morton+al:2025}. It is not surprising to see waves and oscillations in 3D coronal models, as they are fundamental phenomena of the governing equations of the simulations. The question is whether these large scale models (in contrast to dedicated wave models for a single plasma/magnetic loop) have sufficient resolution in time and space to capture the waves on the real Sun. The analysis shown in \citet{Rempel:2017} suggested that the short time scale energy flux contributed by waves might not be a major resource for the coronal energy input compared with the long time scale end. It will be interesting to test whether more energy fluxes related to waves can be generated in higher resolution active region models.

We note that only the disturbance in one temperature range is shown here, but periodic dynamics are commonly seen in plasmas at other temperatures, for example, in loop groups in the lower temperature bins and in the fast upflow region in the hottest bins. In addition to the top view shown in \figref{fig:doppler}, a similar analysis can be performed for the emission measure and Doppler velocity seen from a side view (e.g., along the $x$ or $y$ axis) that mimics the observation of an active region on the solar limb. In that case, the overlapping effect from multiple loop groups along the line-of-sight becomes much more severe, which reduces the contrast of the wave signals to the background. This is very likely an issue that needs to be addressed when measuring the Doppler velocity over the solar limb \citep[e.g., ][]{YangZihao+al:2020,YangZihao+al:2024}. Thus, our model may be a useful test case for assessing such impacts.

\subsection{Conclusion}
To conclude, we present the application of the data-driven MURaM code to construct one-to-one models of observed active regions. The models, which comprise a magnetic evolution stage under the zero-$\beta$ assumption and multiple radiative MHD models for time periods of interest, can capture the emergence of the active region over several solar days and reconstruct the development of the corona of the active region. At the current stage, it is premature to conclude that the numerical models perfectly reproduce every fine structure in the real active region. However, probing the basic magnetic and plasma properties can be performed with parameters chosen on the basis of an educated guess and an affordable computational expense of a few million core hours (including all models presented in this paper). The models quantitatively reproduce the observed EUV intensity within an acceptable range, which can be further improved. We suggest that this method can be applied in more general cases, and the application of the data-driven MURaM code to flare-productive active regions and solar eruptions will be presented in the following paper of this series.

\begin{acknowledgments}
We thank the anonymous referee for many constructive suggestions that improved the clarity and completeness of the paper. F.C. is supported by National Science Foundation of China No. 12422308 and No. 12373054, and by the National Key R\&D Program of China under grant 2021YFA1600504. This work benefits from discussions during the ISSI workgroup ``Data-driven 3D Modeling of Evolving and Eruptive Solar Active Region Coronae". The visualizations shown in \figsref{fig:mag} and \ref{fig:rhot} are created by VAPOR \citep{vapor}.
\end{acknowledgments}

\appendix
\section{Supplementary Animations for Comparing Observed and Model Coronae}\label{sec:appd_animation}
We provide animations that compare observed AIA images side-by-side with model synthesized images. These animations also serve as a complement to the Doppler velocities displayed in \figref{fig:doppler} for inspection of plasma dynamics. All animations are created with the D2\_$\Omega$3 model. Two lines of sight, which are 30$^{\circ}$ and 60$^{\circ}$ away from the vertical direction and parallel to the $y$-$z$ plane, are chosen to calculate the synthetic AIA images in the 131 \AA, 171 \AA,  and 193 \AA~channels. The synthesized images has a cadence of 1000 iterations, which corresponds to 42.6 s. The animations of observed AIA images are generated from www.helioviewer.org.

As discussed in the main text, that the synthesized images yield a systematically lower count rate than the observation does. For the purpose of visual comparison of dynamic evolution, the synthesized images are scaled with lower bounds of dynamical ranges to achieve the best visual similarity to the observed images. In particular, we use a logarithmic scale for the 131 \AA~(1 to 400 DN pixel$^{-1}$ s$^{-1}$) and 193 \AA~(1 and 500 DN pixel$^{-1}$ s$^{-1}$) channels but a square-root scale for the 171 \AA~(10 and 1000 DN pixel$^{-1}$ s$^{-1}$) channel.

\begin{figure*}[t!]
\includegraphics[width=18cm]{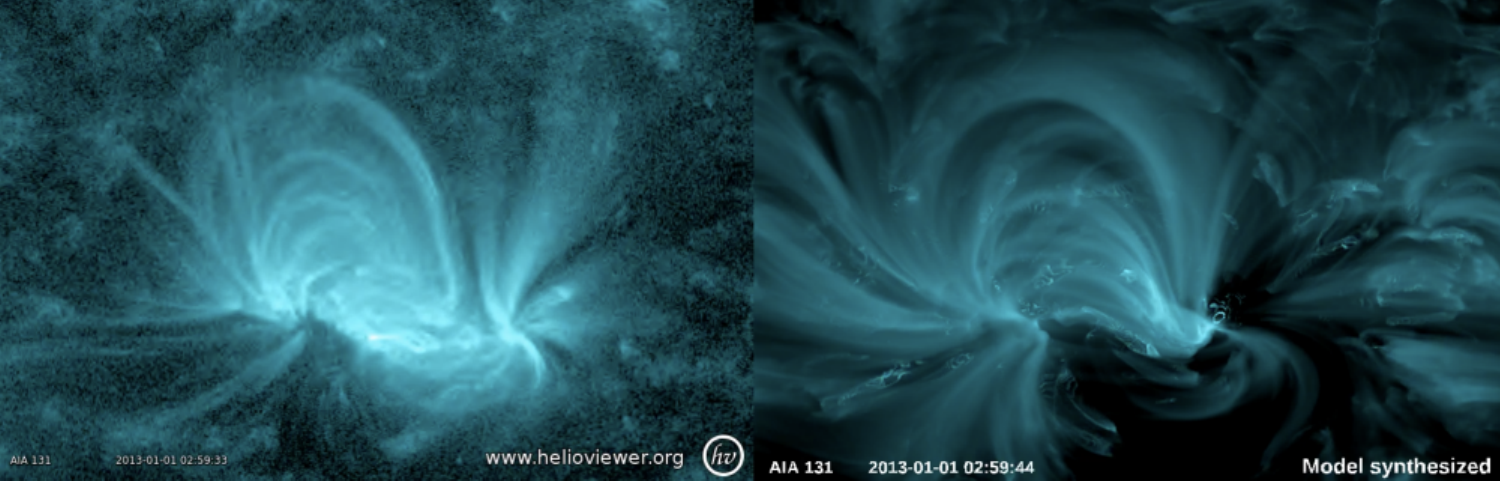}
\caption{A snapshot of an animation that compares the observed and model synthesized AIA 131 \AA~images. The animation, which runs for 17 s, covers the 2.5 hr time period on 2013 January 1 from 00:30:00 UT to 03:00:00 UT. The left half presents an animation of the observed AIA 131 \AA~images. The model synthesized animation (D2\_$\Omega$3 model) is displayed on the right half. The observation animation is generated via www.helioviewer.org. The model synthesized images are calculated through a line-of-sight 30$^{\circ}$ away from the vertical direction and parallel to the $y$-$z$ plane, which is consistent with the latitude of the real active region. The synthesized animation is scaled logarithmically between 1 and 400 DN pixel$^{-1}$ s$^{-1}$, which is chosen to achieve the best visual comparison. The animation is available in the online version of the article.
\label{fig:compare131}}
\end{figure*}

\begin{figure*}
\includegraphics[width=18cm]{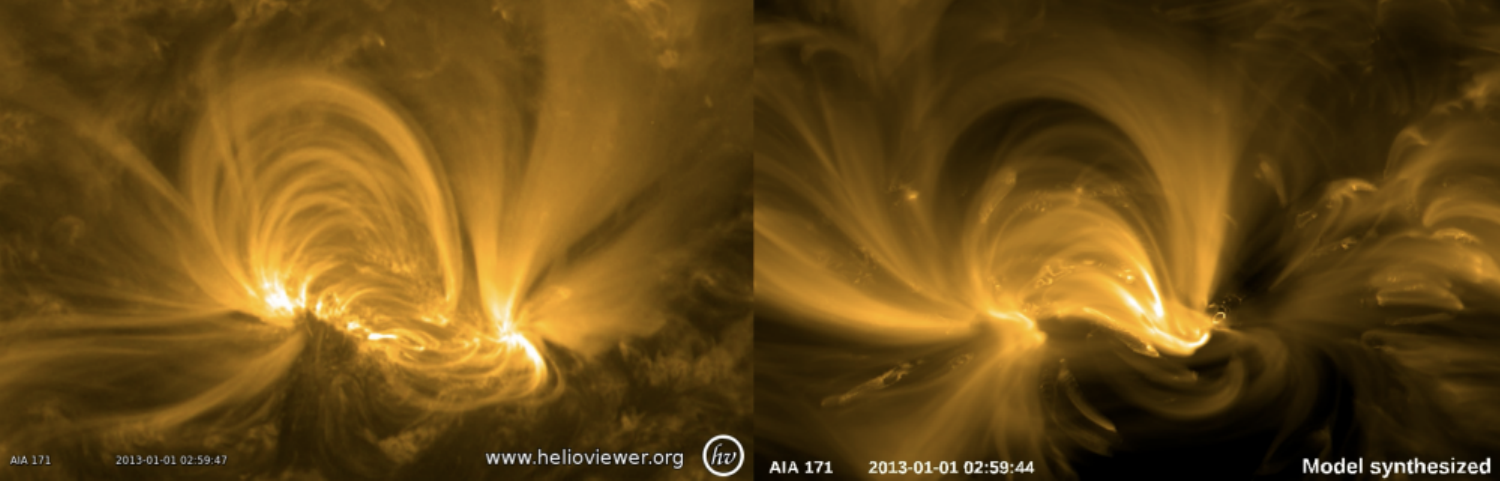}
\caption{A snapshot of animation that compares the observed (left half) and synthesized (right half) AIA 171 \AA~images. The animation follows the same line-of-sight and covers the same time period as \figref{fig:compare131}. The model synthesized animation is displayed on a square-root scale between 10 and 1000 DN pixel$^{-1}$ s$^{-1}$. The animation is available in the online version of the article.
\label{fig:compare171}}
\end{figure*}

\begin{figure*}
\includegraphics[width=18cm]{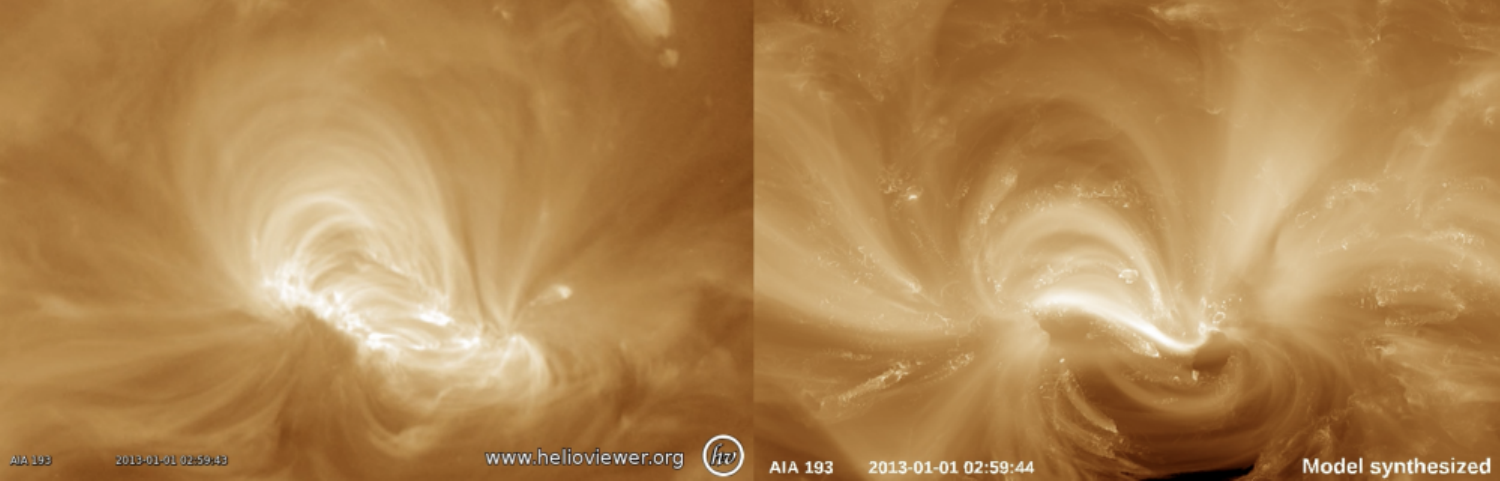}
\caption{A snapshot of animation that compares the observed (left half) and synthesized (right half) AIA 193 \AA~images. The animation follows the same line-of-sight and covers the same time period as \figsref{fig:compare131} and \ref{fig:compare171}. The model synthesized animation is on a logarithmic scale between 1 and 500 DN pixel$^{-1}$ s$^{-1}$. The animation is available in the online version of the article.
\label{fig:compare193}}
\end{figure*}

The animations of synthesized 131 \AA, 171 \AA, and 193 \AA~images are displayed alongside the animations of the observed images in \figsref{fig:compare131}, \ref{fig:compare171}, and \ref{fig:compare193}, respectively. Each animation covers a 2.5 hr time period starting from approximately 2013 January 1 (i.e., Day 2) at 00:30:00 UT.

\begin{figure*}
\center
\includegraphics[width=12cm]{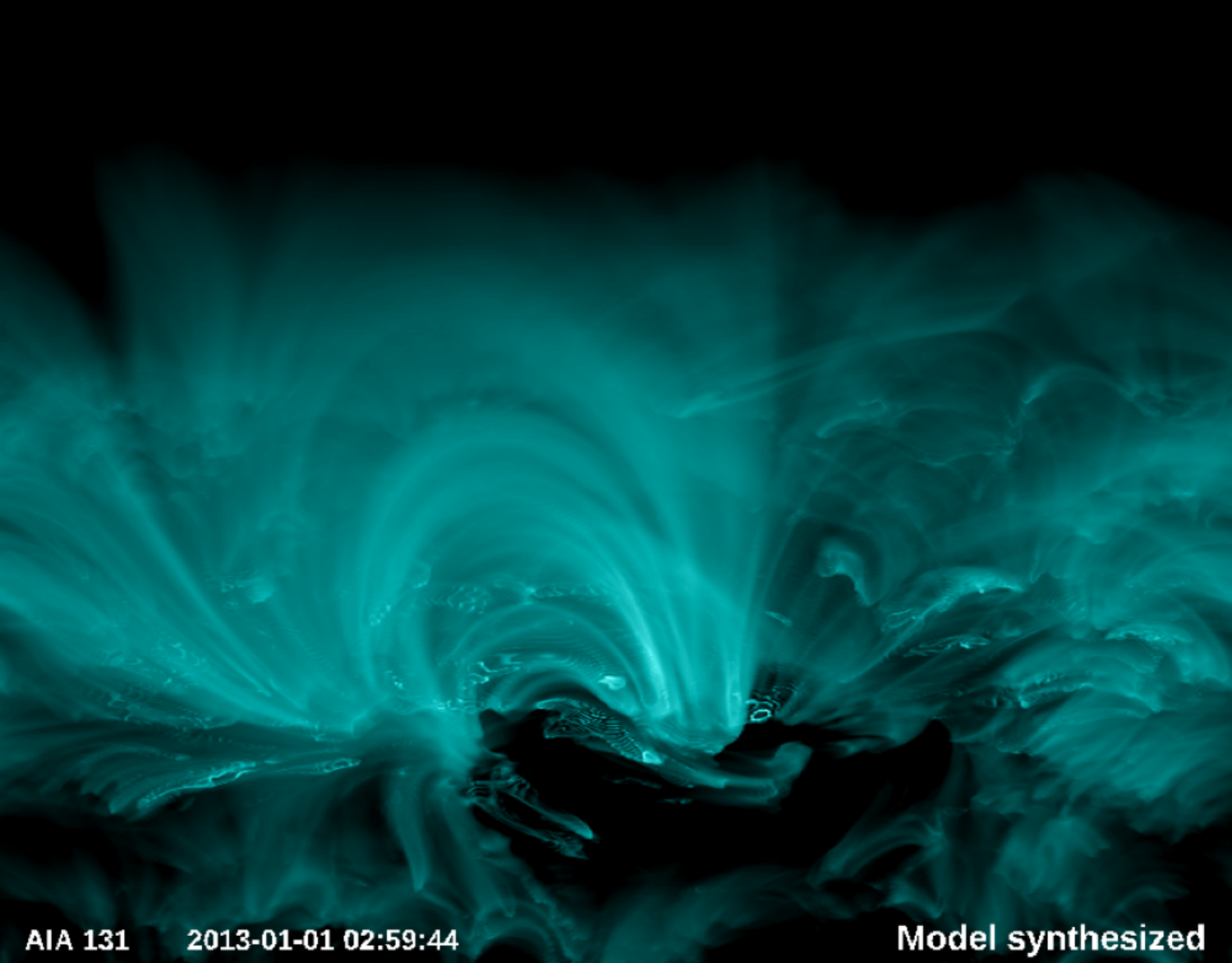}
\caption{A snapshot of an animation that displays the plasma dynamics in the D2\_$\Omega$3 model. The model synthesized AIA 131 \AA~images are calculated along a line-of-sight that is 60$^{\circ}$ away from the vertical direction and parallel to the $y$-$z$ plane. The animation, which runs for 17 s, covers a 2.5 hr time period starting from 2013 January 1 at 00:35:00 UT and is displayed on a logarithmic scale between 1 and 400 DN pixel$^{-1}$ s$^{-1}$. The animation is available in the online version of the article.
\label{fig:los131}}
\end{figure*}

\begin{figure*}
\center
\includegraphics[width=12cm]{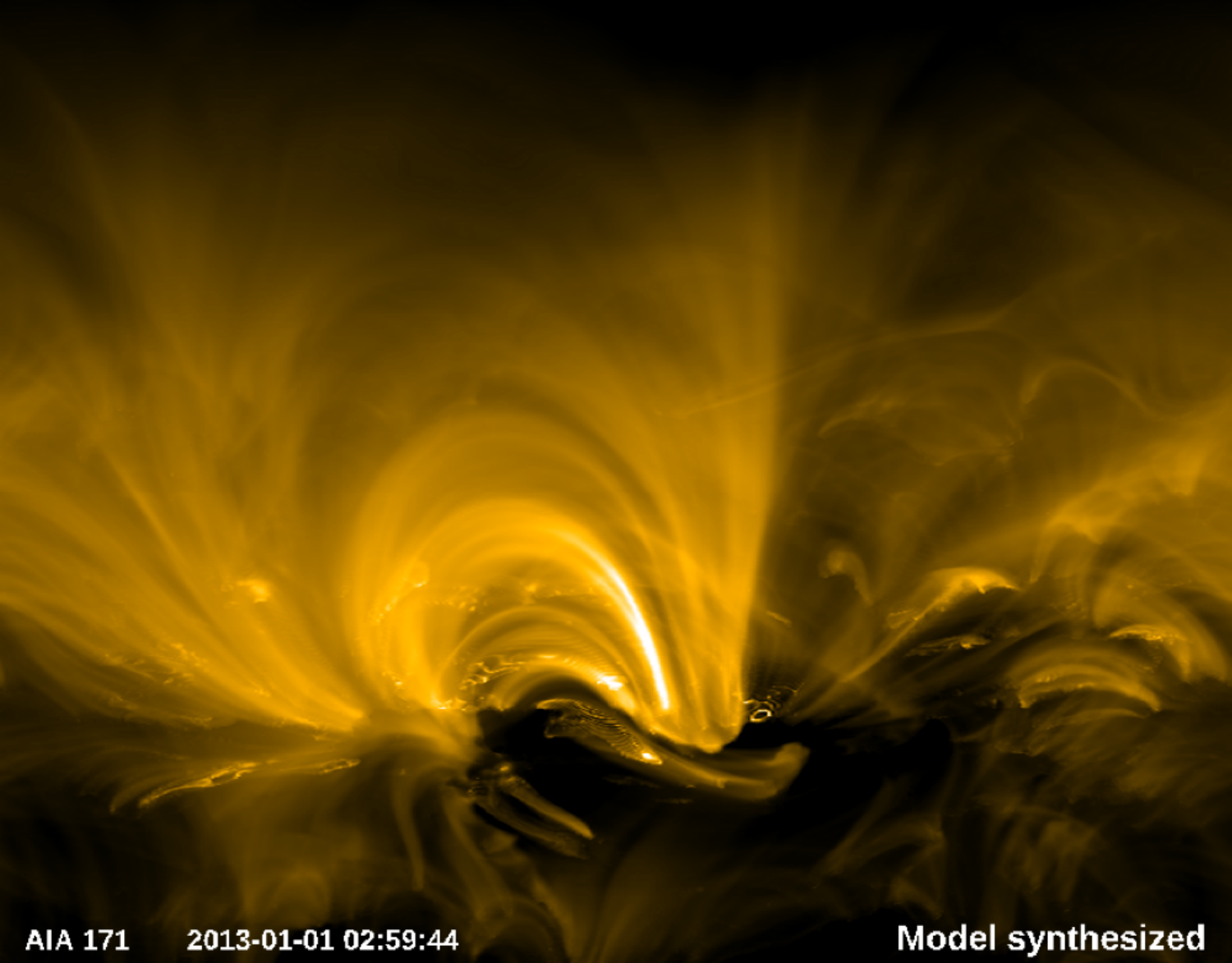}
\caption{A snapshot of an animation of the model synthesized AIA 171 \AA~images following the same method as \figref{fig:los131}. The animation is displayed on a square-root scale  between 10 and 1000 DN pixel$^{-1}$ s$^{-1}$. The animation is available in the online version of the article.
\label{fig:los171}}
\end{figure*}

\begin{figure*}
\center
\includegraphics[width=12cm]{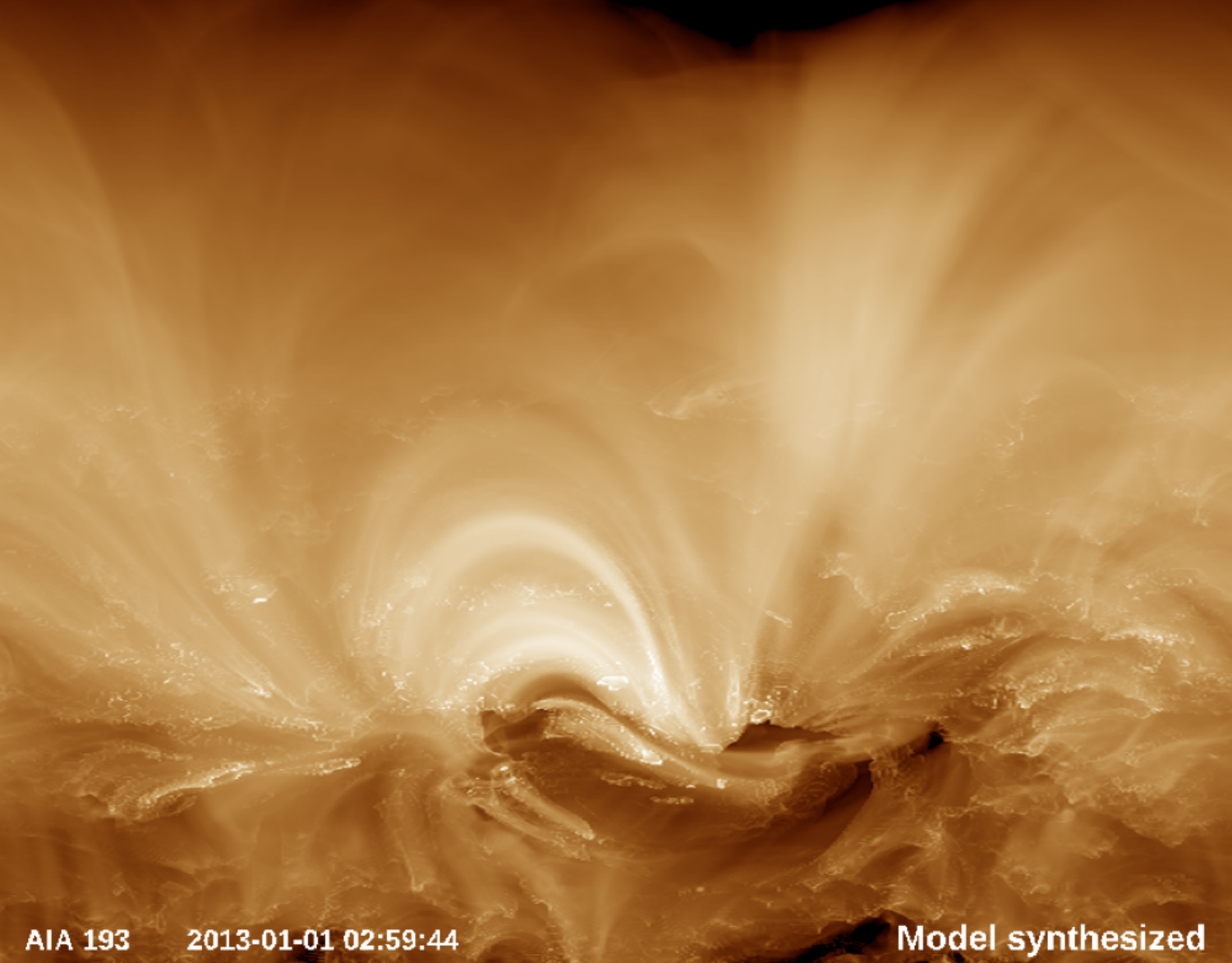}
\caption{A snapshot of an animation of the model synthesized AIA 193 \AA~images following the same fashion as \figsref{fig:los131} and \ref{fig:los171}. The animation is displayed on a logarithmic scale  between 1 and 500 DN pixel$^{-1}$ s$^{-1}$. The animation is available in the online version of the article.
\label{fig:los193}}
\end{figure*}

The animations of the images from the 60$^{\circ}$ view are presented in \figsref{fig:los131}, \ref{fig:los171} and \ref{fig:los193} for the 131 \AA, 171 \AA, and 193 \AA~channels, respectively. The animations cover the same time period and are scaled in the same fashion as \figsref{fig:compare131}, \ref{fig:compare171}, and \ref{fig:compare193}. The view along a more inclined (closer to horizontal) line of sight provides a better illustration of loop-aligned dynamics and those in spicular-like features in the quiet Sun (e.g., top left of the field of view in \figref{fig:los171}). They also demonstrate that the same structure may appear differently with a simple change of the line-of-sight.

\bibliography{reference}{}
\bibliographystyle{aasjournalv7}

\end{document}